**First and corresponding author:** Muhammad Nasir [a, b] (m.nasir@iiu.edu.pk)

**Second author:** Naveed Ikram [b] (naveed.ikram@riphah.edu.pk),

**Third author:** Zakia Jalil [a] (zakia.jalil@iiu.edu.pk)

[a] Faculty of Basic and Applied Sciences International Islamic University, Islamabad, Pakistan

[b] Faculty of Computing, Riphah International University, Islamabad, Pakistan

# Usability Inspection: Novice Crowd Inspectors versus Expert

## Abstract

**Objective:** This research study aims to investigate the use of novice crowd inspectors for usability inspection with respect to time spent and the cost incurred. This study compares the results of the novice crowd usability inspection guided by a single expert's heuristic usability inspection (novice crowd usability inspection henceforth) with the expert heuristic usability inspection.

**Background:** Traditional usability evaluation methods are time-consuming and expensive. Crowdsourcing has emerged as a cost-effective and quick means of software usability evaluation.

**Method:** In this regard, we designed an experiment to evaluate the usability of two websites and a web dashboard.

**Results:** The results of the experiment show that novice crowd usability inspection guided by a single expert's heuristic usability inspection:

a). Finds the same usability issues (w.r.t. content & quantity) as expert heuristic usability inspection.

b). Is cost-effective than expert heuristic usability inspection employing less time duration.

**Conclusion:** Based on the findings of this research study, we can conclude that the novice crowd usability inspection guided by a single expert's heuristic usability inspection and expert heuristic usability inspection, on average, gives the same results in terms of issues identified.

## Research Highlights

- The study investigates crowdsourcing for usability inspection.
- This study compares the results of the novice crowd usability inspection guided by a single expert's heuristic usability inspection with the expert heuristic usability inspection.
- The study shows that novice crowd usability inspection is effective and economical.
- Novice crowd usability inspection is an economical alternative for low-budget organizations.

# 1. INTRODUCTION

Software usability is the degree to which a software system can be used with satisfaction, efficiency, and effectiveness in a particular environment (Abran et al., 2003). There are different types of usability evaluation methods (UEMs), but the prominent two are usability testing and usability inspection. Conventional usability testing is time-consuming and expensive. The real users are not easily available in sufficient number or costly to hire. Moreover, it is difficult to set up an environment to test all aspects of software usability (Nielsen, 1994a; Rosenbaum, 1989; Nielsen, 1989). Usability inspection is an alternative method employing usability experts to evaluate a software system's usability based on their skills, experience, and rules of thumb, also called usability heuristics (Nielsen, 1994a). Usability inspection methods require multiple usability evaluations, i.e., 3-5 expert evaluations (Jeffries and Deservers, 1992), and work well with usability experts only. Although usability inspection methods were introduced as a low-cost means of usability evaluation, there is still a need to find more cost-effective UEMs. Therefore, crowdsourcing has been investigated for usability evaluation to overcome the cost and time delays in traditional UEMs (Sarı et al., 2019).

Crowdsourcing is based on the idea of using the wisdom of the crowd. Surowiecki, in 2005 first introduced the concept of the wisdom of the crowd in his famous book, *The Wisdom of Crowds*. He argued that a large group of people (i.e., crowd) are collectively more intelligent than a couple of experts in problem-solving and decision making. Surowiecki further argued that crowds are not always wise; rather, wise crowds have some common characteristics. The wise crowds should have diverse opinions, and every individual in the group should have an independent view based on their knowledge. Finally, the wise crowd should be able to convert individual opinions into a collective decision.

Crowdsourcing has considerably matured in computing with empirical research (Ambreen and Ikram, 2016), and it has been investigated for usability evaluation (Bruun and Stage, 2015; Guaiani and Muccini, 2015; Liu et al., 2012). Estell' es-Arolas et al. (2012) provided an integrated definition for crowdsourcing. It is a collaborative online activity in which an individual or an organization announces a work task via an open call to a group of people with varying knowledge, skills, and experience on a crowdsourcing platform. The published job can be of variable complexity and completing the task would let the individual earn money, experience, and knowledge. At the same time, the outsourcer will get the job done in return (Estell' es-Arolas et al., 2012).

## 1.1 Research Motivation

Usability engineering has quite rich history spanned over the last couple of decades. However, budget-constrained organizations are often reluctant to adopt usability practices to reduce software development costs. Many studies have identified that one of the major obstacles in adapting usability practices is the budget constraint in small and medium-size software development organizations (Bak et al., 2008; Ardito et al., 2011; Häkli, 2005). In this regard, crowdsourcing has brought new opportunities for usability engineering by overcoming the shortcomings of traditional UEMs (Bruun and Stage, 2015; Guaiani and Muccini, 2015).

Nevertheless, researchers and practitioners face the challenge of crowd workers' poor-quality work (Peer, Vosgerau, & Acquisti, 2014; Goodman, Cryder, & Cheema, 2013). Another problem is the validity of demographic and professional details that crowd worker reports (Liu et al., 2012). Crowdsourced workers sometimes complete tasks multiple times to increase their reward (Kittur, Chi, & Suh, 2008). Therefore, it was worthwhile to explore the wisdom of the novice crowd from the perspective of usability evaluation. One way of exploring the novice crowd's wisdom was to compare usability inspection using crowdsourcing with expert heuristic usability inspection. In this regard, we did not find any research article reporting the comparison between expert heuristic usability inspection and novice crowd usability inspection.

Therefore, there was a need to conduct a research study investigating the novice crowd's wisdom by comparing the novice crowd usability inspection with the expert heuristic inspection. In other words, we wanted to discover the possibility of getting comparable results from novice crowd usability inspection as expert heuristic usability inspection. In this regard, we compared expert heuristic usability inspection with novice crowd usability inspection using crowdsourcing.

We choose expert heuristic usability inspection as a benchmark for comparison as usability inspection methods are economical than traditional usability testing (Nielsen, 1994a; Hollingsed and Novick, 2007). The expert heuristic usability inspection finds more usability problems than any other usability inspection method (Jeffries et al.,1991; Desurvire, Kondziela, & Atwood,1992; Hollingsed and Novick, 2007). Moreover, usability inspection methods can be employed at any development stage of an interactive system (Botella et al., 2013). Nevertheless, some authors support that usability inspection cannot fully substitute usability testing (Hollingsed and Novick, 2007). Besides, some studies have suggested that a combination of end-user and experts-based methods may be employed for more thorough and effective results for usability evaluation (Yen & Bakken, 2009; Hasan et al., 2012). While other authors affirm that there are no real differences between usability testing and inspection methods in terms of usability problems found, and usability testing also misses even critical problems like other usability evaluation methods (Molich and Dumas, 2008). And usability testing is not a gold-standard method to test all other usability evaluation methods (Molich and Dumas, 2008).

## 1.2 Related Work

**1.2.1 Comparative usability studies.** Comparative usability studies have been conducted for the past three decades.

Gray and Salzman (1998) conducted a comprehensive literature survey covering comparative usability studies before 1997. They found that most of the studies suffered from methodological flaws, e.g., (a) low statistical power and wildcard effect that implies that the real differences are not noted, while differences observed are not real, (b) Comparing the outcome of analytical UEMs with empirical UEMs based on the number of usability problems found, instead of the content of usability problems, etc. They also introduced the concept of false alarms, misses, hits, and correct rejections while comparing the UEM's. These flaws challenged the validity of comparative usability studies.

Molich et al. (2004) conducted a comparative usability study by comparing usability testing results conducted by nine independent teams of the same website, i.e., www.hotmail.com. Around 75% of the usability problems were uniquely identified, i.e., no two teams found the same problem. The study observed vast differences in identified usability problems, methodology, tasks, and usability reports submitted by each team. The study concluded that the usability tests' outcome relies highly upon the selected tasks, methodology, and the test controller. Moreover, it concluded that a traditional website might have numerous usability problems, and a typical usability test can only find a small portion of all usability problems. The study emphasized that rather than finding all the usability problems, iterative testing should focus on finding the most critical usability problems.

Molich and Dumas (2008) conducted a comparative usability evaluation study. The study involved 17 teams for the usability evaluation of a hotel's website. Nine teams used usability testing, and eight teams performed an expert evaluation. The study concluded that there are no differences between usability testing and expert reviews in terms of the problems identified, and usability testing is not a benchmark method to compare all other techniques. Usability testing also overlooks problems like any other UEMs, even critical problems. Besides, no evidence was found for the existence of false alarms in expert reviews.

**1.2.2 Comparative usability studies involving novices.** Comparative usability studies involving novices started to appear in the early 90s. Jackob Nielsen performed a comparative study in 1992 to investigate the evaluator effect on heuristic evaluation with varying levels of usability expertise of evaluators, i.e., novice, regular usability experts, and double usability experts. The results showed novice evaluators performed worst, while double experts performed better than regular experts. However, in this case, novice evaluators did not have any experience or knowledge of usability evaluation. Nevertheless, some of the severe usability problems were even found by half the novice evaluators.

Koutsabasis et al., in 2007, conducted a comparative study investigating the performance of novice usability evaluators, having wide background knowledge in interactive design as they graduated in disciplines like graphic design, arts, etc. The study compared nine usability evaluation teams employing different usability evaluation methods, i.e., Heuristic evaluation, Cognitive walkthrough, Think-aloud protocol, and Co-discovery learning. Each group consisted of 3 MSc students studying interaction design as a course. Besides, all the students were familiar with usability evaluation of the websites as part of their coursework previously. The novices first conducted individual evaluations, and later they gathered their findings to interpret the results. The results showed that novice usability evaluators could identify thorough and valid usability problems with some recommendations, i.e., multiple novice usability evaluations must be performed, etc. This study, however, compared the results of the novice usability evaluations with novices.

Law and Hvannberg (2008) conducted an exploratory study to determine how novice evaluators merge and rate the severity of usability problems individually and in collaborative settings. In this regard, the participants were given already collected observational reports of the user testing to examine. The study results show that novice usability evaluation in collaborative settings leads to undesirable effects, i.e., inflation (erroneously increasing severity ratings) and deflation (leniently merging dissimilar usability problems). However, the study compared the performance of novice usability evaluators with novices.

Another study was conducted to explore the effect of using business goals in usability evaluation with novice evaluators (Hornbæk & Frøkjær, 2008). The study results show that novice evaluators employing business goals identified less relevant usability problems than the control group. Moreover, the study compared the results of the novice evaluators with novices.

Howarth et al. (2009) argued that usability evaluation methods do not provide enough guidance to support usability practitioners, especially novice usability evaluators, to perform reliable usability evaluations. They proposed usability problem (UP) instances as a desirable feature to help novices analyze raw usability data (Howarth et al., 2009). Their study shows that usability problem instances can support novice usability evaluators to perform more reliable usability evaluations. However, the study compared the results of novice usability evaluators with novices.

Følstad et al., in 2010, compared the work-domain experts with usability experts in terms of validity and thoroughness for usability evaluation. All the fifteen domain-experts were Ph.D. students with less than a year of experience in the relevant domain; none of them studied HCI as a course or had any experience with usability evaluation previously. The twelve usability experts were hired from different IT companies working as interaction designers or usability consultants. Both usability experts and domain-experts did not have any previous experience of using the test objects. However, domain-experts had domain knowledge and extensive computer experience. The usability inspection method employed was a group-based expert walkthrough. Usability testing was performed as a benchmark to compare results. The study results showed that work-domain experts could produce equally valid but less thorough results for usability inspection compared with usability experts. The study concluded that work-domain experts might be used for usability inspections without compromising validity. Nevertheless, usability inspections with multiple work-domain experts would generate a comparable level of thoroughness as usability experts (Følstad et al., 2010). However, the focus of the study was to investigate usability inspection using work-domain experts, not novice usability inspection.

In their work, Botella et al. (2013) presented a framework involving the concept of design patterns to support novice evaluators performing heuristic evaluations. They proposed that experts' recurring usability problems and relevant solutions to fix them would work as a design pattern to assist novices in their evaluations. However, they did not validate this framework in practical settings.

Borys and Laskowski (2014) investigated whether a large group of novice evaluators can achieve comparable results to a couple of usability experts using heuristic evaluation. The study results rejected the hypothesis that novice usability evaluation results match with expert usability evaluation. However, novice usability evaluators, in this case, had little or no knowledge of usability. All the novice evaluators were computer science students and had attended a training session of an hour to understand the concept of usability evaluation.

De Lima Salgado et al., in 2018, surveyed 38 novice usability evaluators and found that "context of use" is the most challenging usability aspect for novices to understand. The authors suggested that novices may use scenarios, storyboards, and domain-specific principles to conduct heuristic evaluations. Moreover, the authors suggested using collaborative heuristic evaluation for novices, where usability experts analyze usability problem descriptions' quality to shortlist false alarms and duplicate usability problems. However, this study was based on a survey and did not contain any practical comparative evaluation of novice usability evaluations with other usability evaluation methods.

**1.2.3 Design critique-based comparative studies using crowdsourcing.** Some studies have compared experts with novices based on critique/feedback to design UI elements using crowdsourcing. Although design feedback/critique is not the same as usability inspection, the findings of these studies are valuable and relevant to discuss in the perspective of comparing the novice crowd with an expert using crowdsourcing.

Xu et al. (2015) conducted a classroom-based study to investigate the value of feedback received from crowd-workers (non-experts) on posters designed by students as part of an introductory visual design course. The feedback was collected through the structured and free-form method and was compared with experts to judge its quality. The crowd-feedback received was then employed to improve the design of the posters. Later, another iteration of crowd-feedback elicitation was performed to see the effects of earlier changes on posters. The study concludes that the range and depth of expert feedback do not match with crowd-workers. However, crowd-feedback serves as a supplement rather than a replacement for expert feedback.

Moreover, it was found that structured feedback was more interpretive, diverse, and critical than free-form feedback. Nevertheless, crowd workers in this study did not have any design expertise. Besides, this study does not address the challenges of crowdsourcing (discussed in subsection 1.1) and their solution except comparing the quality of the feedback of crowd-workers with experts.

Yuan et al. (2016) conducted a study to investigate whether a rubric of design principles can help novice designers to provide design critique comparable to experts? In this regard, multiple novices and expert designers were recruited from different crowdsourcing platforms. In this experiment, a total of fifteen undergraduate students from a design course were asked to design a dashboard for weather applications. The study results reveal that with the help of expert rubrics, novices can provide a design critique comparable to experts. Besides, this study does not address the other challenges of crowdsourcing (discussed in subsection 1.1) except comparing the quality of the feedback of novice crowd designers (equipped with rubrics of design principles) with experts.

**1.2.4 Usability comparative studies using crowdsourcing.** Usability evaluation studies examining the wisdom of the crowd are few and far between. These studies are discussed in this subsection with their limitations.

Liu et al. (2012) conducted a study on Amazon's Mechanical Turk to compare traditional usability testing with remote usability testing. The object of the experiment was a school's website. In the conventional usability testing part of the experiment, all the participants were experienced users of the website as they were current students at the same school. However, participants in the crowd usability testing were not familiar with the school's website. The results of the crowdsourced usability testing were not as good as traditional usability testing. Nevertheless, crowdsourced usability testing was less expensive, faster, and easy to perform.

In our opinion, the methodology of this study (Liu et al., 2012) could have been improved by controlling some factors. For example, the test group for traditional usability testing was already experienced users of the website that may have influenced the results in favor of conventional usability testing. Moreover, we assume that the crowdsourcing platform selected for this study, i.e., CrowdFlower, also affected results in favor of traditional usability testing as CrowdFlower does not allow to discard crowd workers' poor-quality work. Hence everyone in the crowd usability evaluation received equal payments, despite the quality of the work submitted.

Retelny et al. (2014) introduced the concept of flash teams that can be formulated dynamically using a crowdsourcing platform. The team members in a flash team are experts in a domain that can collaborate to complete a project using a runtime manager named Foundry. One of the tasks in the experiment was the heuristic usability evaluation of a prototype. This study showed that flash teams could complete their work in less time than normal teams on a crowdsourcing platform. However, hiring an expert on a crowdsourcing platform is equally expensive as hiring it conventionally. Besides, this study did not examine the crowd's wisdom for usability evaluation compared to experts.

Bruun and Stage (2015) proposed a usability testing technique called Barefoot. In this technique, Bruun and Stage (2015) suggested that local software practitioners may be given short training to conduct usability testing in resource-constrained organizations. Later, they compared Barefoot with crowdsourced usability testing while hiring university students by sharing minimalist usability training material online. Bruun and Stage (2015) found that the Barefoot approach suits small budget organizations that cannot afford full-time usability experts. Besides that, crowd usability testing is not suitable without HCI competencies. Crowd usability testing requires end-user reports analysis.

In our opinion, the study (Bruun and Stage, 2015) could be improved by controlling the study settings differently. For instance, although asynchronous usability testing (Bruun et al., 2009) is like crowdsourcing in sharing usability evaluation tasks online and completing them remotely (either online or offline) and then submitting them back to the outsourcer online. However, as per the crowdsourcing definition in (Estell' es-Arolas et al., 2012), refers that a work task, i.e., usability evaluation, in this case, may be announced via an open call to a group of people available online having varying skills, knowledge, and experience. We assume that if the sample population for crowdsourcing had been hired through a crowdsourcing platform, results might have differed as crowdsourcing platforms have experienced crowd-workers. Moreover, considering their argument that crowd usability testing does not work well without a crowd's competencies in HCI and usability testing, as valid, we assume that hiring a novice crowd with HCI competencies would not be as expensive as hiring a usability expert. Moreover, this study did not address crowd usability inspection in comparison with expert heuristic usability inspection.

We suggest that designing the tasks (i.e., use-cases) smartly for crowd workers can significantly increase their performance. For example, while developing the use-cases for crowd usability inspection, instead of merely focusing on the system's functionality, use-cases may be designed while keeping in mind the usability guidelines and heuristics to uncover potential usability problems.

**1.2.5 Overview and limitations of related work.** The studies discussed in sub-section 1.2 contain insightful work about comparative usability evaluation, crowdsourcing, or novice usability evaluation. However, these studies have some limitations compared to this study—the study settings of the current literature lack one or more of the following aspects.

1. The study does not involve novices (Gray & Salzman,1998; Molich et al., 2004; Molich & Dumas, 2008).
2. The study does not involve crowdsourcing (Nielsen, 1992; Koutsabasis et al., 2007; Law and Hvannberg, 2008; Howarth et al., 2009; Følstad et al., in 2010; Botella et al., 2013; Borys & Laskowski, 2014; De et al., in 2018).
3. Novice usability evaluators have no experience and knowledge of usability evaluation (Nielsen, 1992; Borys and Laskowski, 2014)
4. Novice usability evaluation has been compared with novices (Koutsabasis et al., 2007; Law and Hvannberg, 2008; Hornbæk & Frøkjær, 2008; Howarth et al., 2009).
5. The study presents a theoretical framework for novice usability evaluation without validation (Botella et al., 2013).
6. The study presents suggestions based on a survey of novice usability evaluators (De et al., in 2018).
7. The study does not address the challenges of crowdsourcing or proposed any solution for it (Liu et al., 2012; Bruun & Stage, 2015; Xu et al., 2015; Yuan et al., 2016).
8. The studies are based on the design critique of UI elements but do not specifically compare usability inspection methods (Xu et al., 2015; Yuan et al., 2016).
9. The studies support novices with features different from our proposed study, i.e., UP instances, domain knowledge, a rubric of design principles, short usability training, etc. (Howarth et al., 2009; Folstad et al., 2010; Botella et al., 2013; Bruun & Stage, 2015).
10. The study settings could have been improved in some cases, e.g., the control group and the treated group could have the same learning experience with test objects to avoid any biases, etc. (Liu et al., 2012).

We will discuss the overview of each subsection of related work with their limitations here further. The studies discussed in subsection 1.2.1 brought valuable insights regarding methodological flaws and misconceptions while comparing usability evaluation methods. However, these studies were not focused on novices.

Subsection 1.2.2 discusses comparative usability studies involving novices. However, these studies were conducted in traditional settings and did not involve crowdsourcing. Besides, the study conducted by Jackob Nielsen in 1992 involved novices that did not have any experience or knowledge of usability evaluation. It has been established in several studies that novices that do not have any practical experience or knowledge of usability evaluation would not be able to perform significantly in usability evaluation (Borys and Laskowski, 2014). Another limitation of most of the studies discussed in subsection 1.2.2 is that they compared the performance of novice usability evaluators with novices (Koutsabasis et al., 2007; Law and Hvannberg, 2008; Hornbæk & Frøkjær, 2008; Howarth et al., 2009). Other studies included in subsection 1.2.2 either proposed a theoretical framework for novice crowd-workers without validation (Botella et al., 2013) or presented suggestions based on a survey (De et al., in 2018).

Subsection 1.2.3 discusses studies that compared the design critique of novice crowd-workers/designers in comparison with experts involving crowdsourcing. However, these studies did not compare usability evaluation methods (Xu et al., 2015, Yuan et al., 2016). Besides, these studies did not address the challenges of crowdsourcing except the quality of the feedback of crowd-workers.

Subsection 1.2.4 is focused on comparative usability studies involving crowdsourcing. The study conducted by Liu et al. (2012) concluded that remote usability testing employing crowdsourcing did not perform well compared to traditional usability testing. However, the participants in conventional usability testing were already experienced users of the test object, while this was not the case with crowd usability testers. In another study by Retelny et al. (2014) suggested that a group of experts can work more productively using crowdsourcing than traditional teams. However, this study did not involve novice crowd usability evaluators, neither they addressed the challenges of crowdsourcing. Bruun and Stage (2015) concluded that crowd usability testing does not perform well without skills and experience in HCI. However, in our opinion, hiring novice crowd inspectors with HCI skills and experience are not as expensive as usability experts. Moreover, these studies did not address the challenges of crowdsourcing except the quality of the work of crowd-workers.

Besides, we have not found any other study comparing novice usability inspectors with experts involving crowdsourcing. Most of the usability studies (discussed in subsection 1.1) involving crowdsourcing do not address the challenges of crowdsourcing nor provide any reasonable solutions for these challenges to the best of our knowledge except comparing results of crowd-workers with control groups.

To overcome the limitations of the existing literature discussed earlier, this study investigates novice usability inspection in comparison with expert heuristic usability inspection using crowdsourcing while addressing the challenges of crowdsourcing and proposing a reasonable solution for these challenges.

The rest of the paper is organized as follows: section 2 discusses Research Method, section 3 is Analysis and Discussion, section 4 is Limitations and Validity Threats. Section 5 is the Conclusion and Future Work.

## 2. RESEARCH METHOD

Our chosen research method is experimentation in this research study since we want to study the cause-and-effect relationship between expert heuristic usability inspection and novice crowd usability inspection with use-cases. The complete experiment design is illustrated in Figure 1.

### 2.1 Goal and Definition

The goal statement of this experiment is as below:

"*Analyzing Crowdsourcing for the purpose of Usability Inspection with respect to Wisdom of crowd from the point of view of Researcher using two websites and a web dashboard.*"

The artifacts of the experiment definition are as below:

**The object of study.** The object of the study is crowdsourcing. In this regard, we explored different crowdsourcing platforms like Amazons Mechanical Turk (MTurk), uTest, and Upwork. We found that MTurk does not work outside the US due to the payments and tax issues (MTurk, 2015). Moreover, MTurk allows crowd-workers to submit their work anonymously (mTurk, 2020). Therefore, it is impossible to generalize the findings to a target population (Stritch et al., 2017). Besides, MTurk allows outsourcing work in microtasks which is not suitable for large assignments such as usability inspections.

Besides exploring uTest quite a bit, we were not able to find how to post our project on uTest (uTest, 2016). Other people have also reported similar issues with uTest (glassdoor, 2021). Finally, we selected Upwork, as it was easy to learn and simple to use (Upwork, 2020).

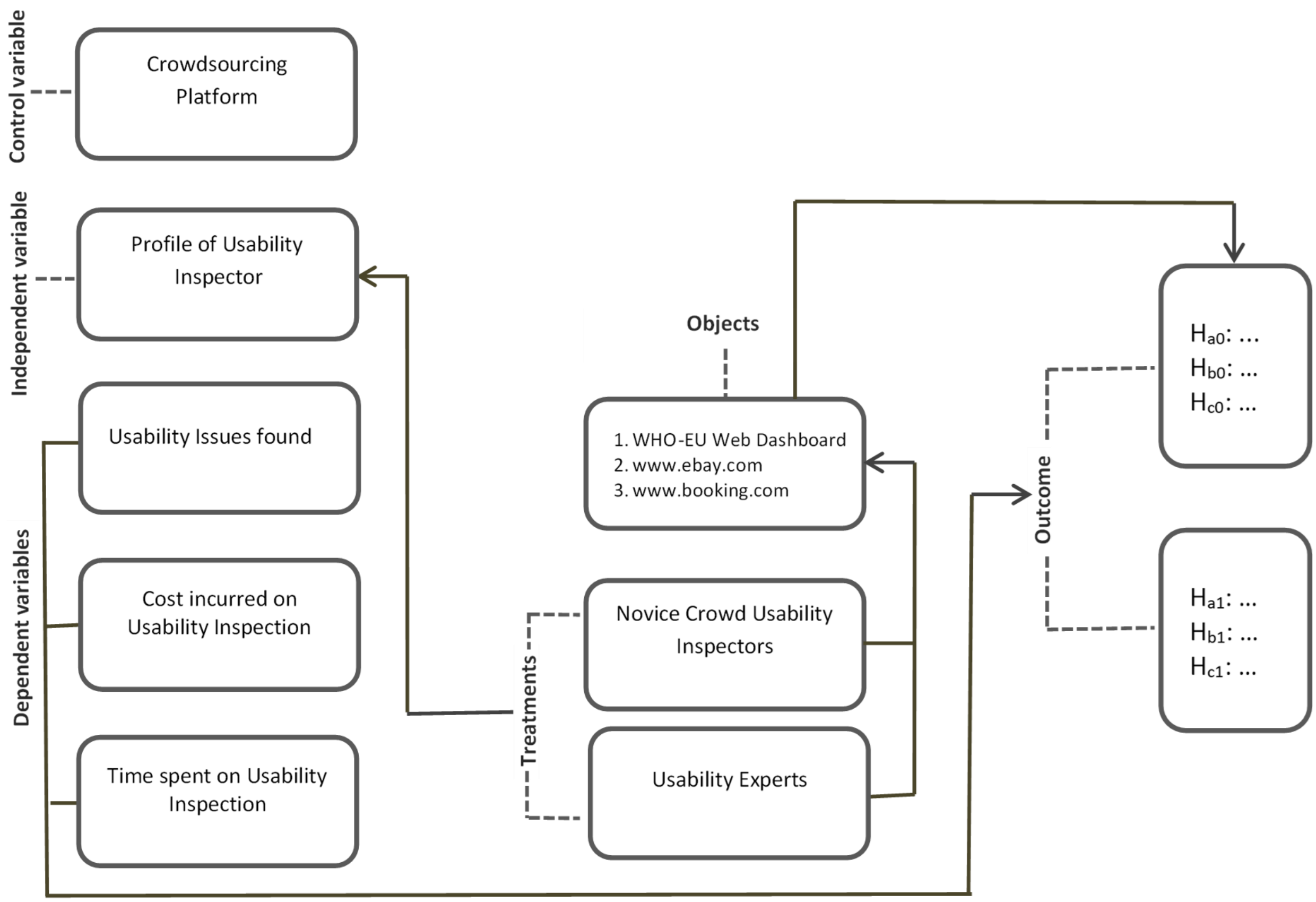

**Figure 1: Experiment Design**

**Purpose.** The purpose of this experiment is to compare the expert heuristic usability inspection with novice crowd usability inspection.

**Quality Focus.** The quality focus of this experiment is to evaluate the wisdom of the crowd. We want to see whether novice crowd usability inspection can give us the results that are equally good as expert heuristic usability inspection or not? Other quality foci are time and cost incurred on both usability inspection methods.

**Perspective.** We are conducting this experiment from the perspective of the researcher.

**Context.** This experiment is conducted using a crowdsourcing platform, i.e., Upwork. Usability experts, as well as novice crowd inspectors, were hired using Upwork. Details about the test objects and subjects are discussed in subsection 2.2.3 instrumentation.

## 2.2 Planning

The experiment's planning stage consists of hypothesis formulation, variable selection, subject selection, design selection, instrumentation, and validation.

**2.2.1 Hypothesis Formulation.** Following hypotheses have been formulated for this experiment:

$H_{a0}$ – Novice crowd usability inspection on average finds different usability issues (w.r.t. content & quantity) compared to expert heuristic usability inspection.
$H_{a1}$ – Novice crowd usability inspection on average finds the same usability issues (w.r.t. content & quantity) as expert heuristic usability inspection.

$H_{b0}$ – Novice crowd usability inspection incurs the same cost as expert heuristic usability inspection.
$H_{b1}$ – Novice crowd usability inspection incurs less cost than expert heuristic usability inspection.

$H_{c0}$ – Novice crowd usability inspection, on average, takes the same time as expert heuristic usability inspection.
$H_{c1}$ – Novice crowd usability inspection, on average, takes less time than expert heuristic usability inspection.

**2.2.2 Variables and Subjects' Selection.** There is only one independent variable in this experiment, i.e., the profile of the usability inspector. There are two treatments for it, i.e., usability experts and novice crowd usability inspectors. The dependent variable that we are measuring are usability issues found, cost incurred, and time spent on usability evaluations, as illustrated in Figure 1. We blocked the effect of two control variables in this experiment, i.e., the crowdsourcing platform and the evaluator effect. We used the same crowdsourcing platform, i.e., Upwork, for the whole experiment to block its effect. Moreover, to block the evaluator effect, we hired multiple usability evaluators, i.e., at least five usability inspectors, for each experiment trial for both expert heuristic evaluation and novice crowd usability inspection (Hertzum & Jacobsen, 2001). Details about the usability experts and novice crowd inspectors will be discussed in subsequent subsection 2.2.3 Instrumentation.

**2.2.3 Instrumentation.** There are two parts to the experiment. Therefore, we designed separate instruments for each section. However, to compare the UEMs, we need a standard coding scheme to identify and classify usability problems.

**Coding Scheme:**

The coding scheme used in this comparative study, provided in Appendix E, is derived from Molich and Dumas's (2008) work. The focus of the coding scheme is to transform raw usability descriptions into usability problems. In this regard, the coding scheme has classified raw usability descriptions into different artifacts, i.e., atomic comment, problem comment, false positive, minor problem, serious problem, critical problem, issue, key issue, etc. The coding scheme's application with a complete data analysis method is described in subsection 3.1, Analyzing usability inspection reports.

**Test Objects:**
In the first trial of this experiment, we evaluated the usability of a web dashboard of the World Health Organization designed for Europe (WHO EU Health e-Atlas). We chose a web dashboard as it contains a variety of complex visual elements. A web dashboard plays a significant role in data mining (Kostkova et al., 2014). A web dashboard aims to present large data in a concise way to increase its understanding. A well-designed dashboard helps in decision making and allows users to see the data from different perspectives and co-relate different factors in it (Kostkova et al., 2014). The significance of web dashboards in the public health sector is even more vital as they can help us monitor disease outbreaks (Lechner and Fruhling, 2014). Therefore, we chose the WHO-EU web dashboard (WHO-EU, 2016) designed to provide information about core health statistics, including demographics, health status, risk factors, health resources and their utilization, and expenditure in 53 countries of Europe.

We diversified the same experiment with two more trials to add more confidence in results, using two general-use websites from different domains, i.e., www.ebay.com and www.booking.com. We chose these two websites as these are typical websites of public use. eBay.com is a website for online shopping and auction with business operations in 30 countries. The website does not charge anything to buyers, but sellers are charged for listing items after a couple of free listings and selling items. Booking.com provides travel and accommodation services in 43 different languages. The website has more than 28 million listings for accommodations to stay. The weblinks for all three test objects are provided in Appendix A.

**Expert Usability Inspection's Instrument:**
A criterion was defined for the selection of usability experts. A typical usability expert must have:

- Master's or Ph.D. degree in HCI or relevant field
- At least five years' practical experience in usability evaluations
- Preferably be a certified usability expert, i.e., HFI CUA (Human Factors International Certified Usability Expert)

Weblinks for the test objects were shared with usability experts for expert heuristic evaluation, see Appendix A.

**a)** ***Usability Heuristics.*** A list of usability heuristics was shared with usability experts; see Appendix B. The heuristics for the WHO-EU web dashboard were focused on usability inspection of the web dashboard and information visualization. These heuristics were previously validated in an MS thesis study (Ghaffar & Nasir, 2016). Ghaffar & Nasir (2016) examined the existing literature comprehensively to collect all the heuristics that have been reported for designing web-dashboard and information visualization. The authors divided these heuristics into two sets, i.e., common heuristics (set 1) and common plus other heuristics (set 2). The common heuristics were grouped using Nielson's (1994b) and Schneiderman's (1996) heuristics. The heuristics that did not match Neilson's and Schneider's heuristics were considered in the category of other heuristics. The study's objective was to find out the most suitable set of heuristics for designing web dashboards and information visualization. An experiment was conducted to compare the effectiveness of both sets of heuristics. In this experiment, four groups of students, each containing one undergraduate student of $8^{th}$ semester from Bachelor of Software Engineering, were randomly selected. Group A was treated with heuristic set 1, while group B was given heuristic set 2 to design web dashboards to represent information for POLIO vaccination in Pakistan. The usability testing of the developed web dashboards revealed that the web dashboard designed with heuristic set 2 (common+other) were better in usability.

The rest of the two test objects were evaluated using ten usability heuristics for user interface design by Jakob Nielsen, see Annex A (Nielsen, 1994b).

**b)** ***Payment Criterion***. Usability experts were offered a fixed price contract based on their previous experience, skills, and qualification for performing heuristic usability inspections on Upwork. In this regard, we thoroughly checked their profiles on Upwork, including information like jobs completed previously on Upwork, payments, and feedback received to verify their skills and per hour rate.

**c)** ***Time Management:*** Each expert was requested to record the time it took to perform the evaluation.

**d)** ***Roles:*** Experts were asked to evaluate the test objects with the following roles:

a) Buyer for eBay.com
b) Guest/Tourist looking to book accommodation for Booking.com.
c) Any visitor seeking health-related information and trends for WHO-EU web dashboard.

**e)** ***Format for Evaluation:*** Experts were asked to document usability problems with at least the following information.

- Title for the problem
- Brief description of the problem
- Heuristic(s) violated.
- Type: Minor problem, Serious problem, Critical problem
- Snapshot of the area of UI where the problem was found.

**Novice Crowd Usability Inspection's Instrument:**

The following criterion was used for the selection of novice crowd usability inspectors. As per this criterion, a typical novice crowd usability inspector preferably has:

- A bachelor's degree/diploma in computing or relevant field/attended a course relevant to HCI.
- 1-2 years of experience in software usability evaluation/software quality assurance

testing/software testing

Botella et al. (2014) proposed a classification scheme for usability evaluators based on experience and university degrees. They suggested that a novice usability evaluator has a university degree, attending one course related to HCI with at least a few hours of usability evaluation experience.

However, it has been established in several usability studies that novice evaluators with no or few hours of usability evaluation training/experience cannot perform comparably to usability experts (Nielsen, 1992, Borys & Laskowski, 2014). On the other hand, the studies that employed novices with extensive computer experience and background knowledge in usability evaluation produced more encouraging results than usability experts (Koutsabasis et al., 2007, Howarth et al., 2009, Folstad et al., 2010).

Therefore, motivated by the existing literature, we set the criterion for novices to preferably have attended a course in HCI and at least 1-2 years of practical experience in usability evaluation/software quality assurance/software testing. However, Botella et al. (2014) classified a person without a university degree attending several courses on HCI or with less than 2500 hours of practical experience in usability evaluation as a beginner. On the other hand, studies that employed novices with less than a year's practical experience in usability evaluation with a course in HCI also referred to them as novices (Howarth et al., 2009). Therefore, the boundary between the novice and beginner is overlapping. Either we call it a more experienced novice or beginner, we considered the university degree/diploma/certification in usability evaluation/HCI, besides the practical usability experience of the evaluators, as an independent variable in the design of our study. Therefore, we referred to it as the profile of the usability inspector, see Figure 1. Moreover, we used the term novice for the selection criterion set earlier in this study for novice crowd usability inspectors.

The artifacts that were shared with novice crowd usability inspectors to perform usability inspection are as below:

***a)*** ***Usability Guidelines***. The guidelines were provided to identify usability problems in the test objects. The guidelines contained a list of usability heuristics, the same as shared in expert usability evaluation, besides instructions about how to attempt questionnaires.

***b)*** ***Questionnaires for Novice Crowd Usability Inspection***. The existing evidence on novice usability evaluation supports the argument that without assisting tools (expert rubrics, design patterns, storyboards, domain-specific principles), novice usability inspectors may not be able to perform comparably to usability experts (De et al., 2018; Borys & Laskowski, 2014, Botella et al., 2013, Hornbæk & Frøkjær, 2008). Therefore, the authors designed the questionnaires for each test object to support novice inspectors. Each questionnaire contained 20 Use-Cases. The questionnaires were based on the usability inspection report of a single expert for each test object. However, not all the use-cases were based on the expert heuristic evaluation. Instead, half of the total use-cases were general use-cases based on the different functionalities of the system. Moreover, expert heuristic usability inspection involved five usability experts, while the questionnaire for novice inspectors was based on a single expert's usability inspection report.

Therefore, the questionnaire for novice usability inspection may not entirely depend on the expert's findings and, consequently, the results.

Other studies have also suggested supporting novices with usability experts or artifacts extracted from expert's usability inspections reports (Botella et al., 2013; De et al., 2018). For example, Botella et al. (2013) proposed a framework based on the recurring usability problems identified by usability experts to assist novices. Another study by De et al. (2018) suggested collaborative usability inspection where the novices may be supported with usability experts to analyze the quality of usability problems identified by novices, identifying false alarms and duplicate usability problems. They also suggested supporting novices with scenarios, storyboards, and domain-specific principles.

The WHO-EU web dashboard questionnaire contained a couple of prerequisite tasks for novice crowd inspectors to familiarize themselves. We did not include any prerequisite tasks for eBay.com and Booking.com as these test objects were easy to use. After performing each use case, novice crowd inspectors were supposed to provide feedback regarding any usability problems found.

The questionnaire for the first trial of novice crowd usability inspection was developed using Google Forms. However, for the rest of the two trials of novice crowd usability inspection, questionnaires were simply designed in MS Word, as Google Forms does not save a questionnaire before submission. As a result, the user must complete it in one sitting. The weblinks for questionnaires are provided in Appendix C.

The steps to design questionnaires are listed below:

1. Extract usability problems from an expert heuristic evaluation using the coding scheme provided in Appendix E. Detailed process to transform usability descriptions into usability problems is discussed in section 3.1, Analyzing Usability Inspection Reports.
2. Take a screenshot of the UI area where the usability problem was reported by an expert and mark that area on the screenshot.
3. Develop a general use case to guide the novice crowd inspector to execute a task relevant to the usability problem.
4. After performing the task, the crowd inspector shall be requested to document any usability problem found in the space provided below the use case.

The following example of a use-case from eBay.com explains how a usability problem comment can be employed to develop a use-case for a questionnaire.

**Example:**

***Usability problem comment:***

*A user is not able to select multiple items within one refiner category (E.g., Select AT&T and T-Mobile under Network). It is a UX glitch that hinders the refiner experience.*

***Use Case:*** *Select multiple choices from side filters. E.g., For a mobile phone, you can choose Apple, LG, etc. Were you able to select multiple categories in filters easily?*

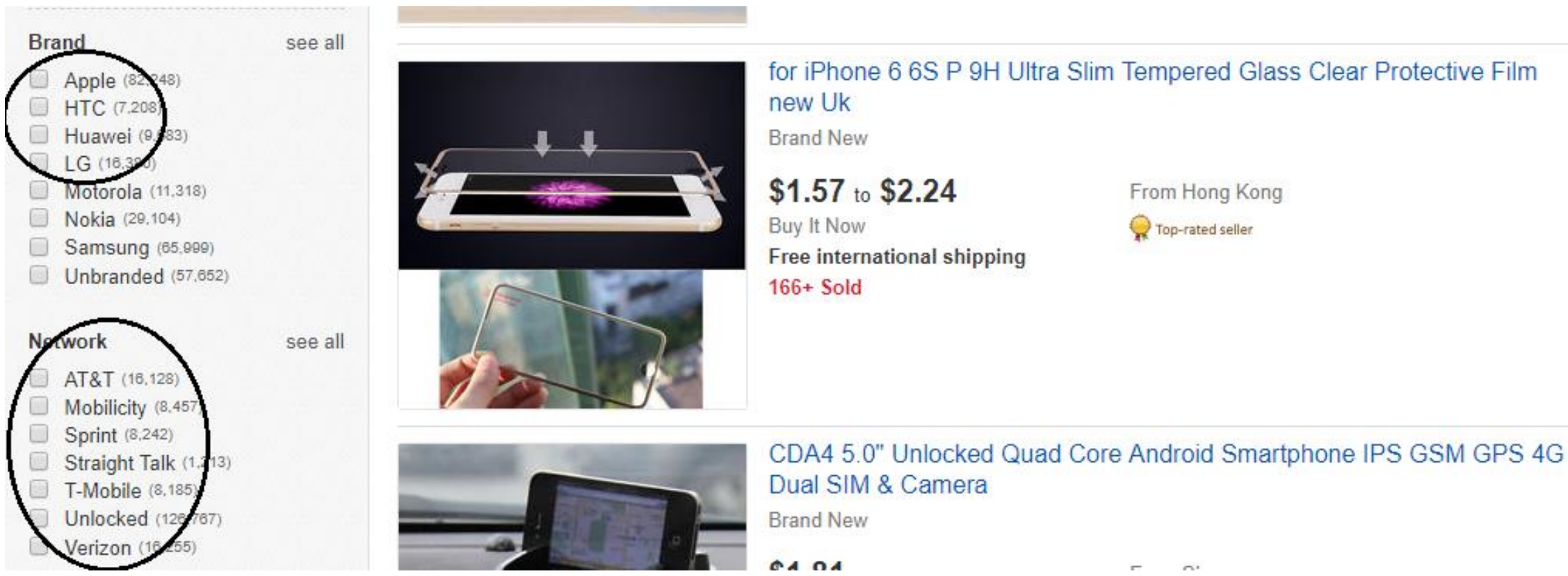

After performing this task, if you found any usability problem(s), please document them below:

It has been already validated in research that without assisting tools (expert rubrics, design patterns, storyboards, domain-specific principles), novice usability inspectors cannot perform to a level comparable to usability experts (De Lima Salgado et al., in 2018; Borys and Laskowski, 2014, Botella et al., 2013, Hornbæk & Frøkjær, 2008). Therefore, we did not need to validate the questionnaires employing an independent group of novice crowd inspectors that were not treated with the questionnaire. Hence, we considered the questionnaire part of the novice crowd usability inspectors' profile instead of a separate intervening variable.

***c)*** ***Payment Criterion***. We set the payment criterion for novice crowd usability inspection based on performance as below:

- A novice crowd inspector would earn 1 USD for the identification of 3 valid usability issues.
- A novice crowd inspector can earn a maximum of 10 USD to identify 30 valid usability issues.

***d)*** ***Roles and Time Management***. Crowd inspectors were asked to evaluate test objects with the same roles as experts. Besides, crowd inspectors were requested to record time for questionnaire-based usability evaluation.

### 2.3 Execution of Usability Evaluations

We conducted heuristic usability inspection first since we were planning to use the results of one of the five expert's heuristic evaluations to design questionnaires for novice crowd usability inspection. All the evaluators, including experts and crowd usability inspectors, were asked to submit their English language evaluations. The authors ensured that the test objects did not change through the duration of expert and crowd evaluations. Consequently, evaluators may get the same test objects to evaluate. Time durations for usability evaluations were as below:

a. WHO-EU web dashboard – January 2016
b. www.ebay.com – February 2019
c. www.booking.com – February 2019

#### 2.3.1 Heuristic Usability Inspection

In expert heuristic usability inspection, a total of 8 usability experts were self-selected based on a set criterion, discussed in subsection 2.2.3 instrumentation. Each usability expert was assigned an ID starting from E1 to E8. Two experts, i.e., E1 and E2, participated in all three experiment trials, while experts for trials 2 and 3 were the same, as shown in Table 2. A single expert performed each heuristic usability inspection.

***Qualification:*** Three experts (E1, E2, and E3) had certification in usability evaluation, one expert (E5) had a Ph.D. in cognitive science, one expert (E6) had a master's in Psychology and four experts (E3, E4, E7, and E8) had master's degree in HCI.

***Experience:*** Usability experts had an average practical usability evaluation experience of 7.2 years, with a minimum of 5 years and a maximum of 15 years (see Table 1).

***Payments:*** The fixed-price contracts ranged from USD 200 to USD 400, with an average of USD 280 per expert heuristic usability inspection (see Table 4).

***Diversity in hiring:*** Experts were invited from worldwide, and the final selection included experts from Asia, Europe, North and South America.

***Familiarity with test objects:*** None of the experts was familiar with the WHO-EU web dashboard before the experiment. However, expert E1 hardly used eBay.com and used Booking.com only a couple of times. Expert E2 visited eBay.com a couple of years ago before evaluation but never bought anything. Besides, Expert E2 never used Booking.com; instead used a similar website Blocket.se that is commonly used in Sweden for booking accommodation. Expert E6, E7, and E8 visited these websites once in a blue moon before evaluation and were not frequent users of these websites.

***English language skills:*** All the experts were non-native English speakers except E5, E6, and E7. However, none of the experts had any communication problems in the English language during evaluation or communication with the authors.
Each expert performed a heuristic evaluation independently and submitted evaluation reports to the authors for further analysis.

### 2.3.2 Novice Crowd Usability Inspection

A total of 11 novice crowd usability inspectors were self-selected based on a set criterion discussed above in subsection 2.2.3 instrumentation. We assigned each novice crowd usability inspector an ID starting from C1 to C11. Five novice crowd inspectors (C1 to C5) participated in the first trial of the experiment. The other six novice crowd inspectors (C6 to C11) participated in the experiment's second and third trials, as shown in Table 2. A single novice crowd usability inspector performed each inspection.

***Qualification:*** A total of 10 out of 11 novice crowd inspectors had a bachelor's degree, i.e., three in Computer Sciences (C3, C9, and 10), two in IT (C2, C7), and one in each, Computer Applications (C4), Software Engineering (C11), Informatics (C5), Arts (C6), and in Business Administration (C8). Only one novice inspector had a Diploma in Computer Science (C1).
***Experience:*** All novice crowd usability inspectors, except C8, were familiar with HCI/usability/quality assurance testing. Besides, each novice crowd inspector had 1-2 years of experience in software testing except C3 (see Table 2).

***Payments:*** The payment for each novice crowd inspector ranged from USD 1 to USD 10, with an average amount of USD 6, see Table 4.

***Diversity in hiring:*** Novice crowd usability inspectors were hired on Upwork from different parts of the world, including Russia, India, UK, Turkey, Pakistan, Armenia, Ukraine, Kenya, Norway see Table 2.

***Familiarity with test objects:*** In the 1st trial of the experiment, none of the novice crowd inspectors was familiar with the test object, i.e., the WHO-EU web dashboard. In the 2nd and 3rd trials of the experiment, except for C8, all the novice crowd inspectors (C6, C7, C9, C10, and C11) were familiar with at least similar test objects (see Table 2).

***English Language Skills:*** All the crowd inspectors were non-native English speakers, except C8 (see Table 2). The usability comments of crowd inspectors were understandable primarily. However, if any confusion was observed in the language of usability comments, we clarified them from crowd inspectors using Upwork chat.

***Validation of the crowd's identified usability problems:*** Authors validated the crowd's identified usability problems by comparing them with expert issues identified. Only 23 usability problems did not match with experts' identified usability problems. For further validation, we sent these 23 usability problems to usability expert E1; see section 3.1 Analyzing usability inspection reports for more details.

Crowd workers often submit poor quality, incomplete, and fake work to claim payments. Gomide et al. (2014) suggested that the work of outliers (spammers) should be filtered out. However, we did not filter out any usability inspection reports to give a more objective view of the results.

**Table 1. Profiles of Usability Experts**

| Experiment I – WHO-EU Web Dashboard | | | | | | |
|---|---|---|---|---|---|---|
| **Resource ID** | Country | Education | Certification in Usability | Academic Experience (Years) | Practical Usability Experience (Years) | Currently Working as |
| **E1** | India | MSc in Computing | HFI-CUA | - | 6.5 | Senior UX Designer |
| **E2** | Sweden | MSc in Cognitive Science | HFI-CUA/CXA | 5 | 15 | UX/Usability Expert |
| **E3** | Brazil | MSc in HCI | Nielsen Norman Group UX Researcher | - | 6 | UX/Usability Designer |
| **E4** | Germany | MSc in HCI | - | - | 5 | UX/Usability Consultant |
| **E5** | America | Ph.D. in Cognitive Science | - | 4 | 8 | Senior UX Researcher |
| **Experiment II – www.ebay.com & Experiment III – www.booking.com** | | | | | | |
| **Resource ID** | Country | Education | Certification in Usability | Academic Experience (Years) | Practical Usability Experience (Years) | Currently Working as |
| **E1** | India | MSc in Computing | HFI-CUA | - | 6.5 | Senior UX Designer |
| **E2** | Sweden | MSc in Cognitive Science | HFI-CUA/CXA | 5 | 15 | UX/Usability Expert |
| **E6** | USA | MSc in Psychology | - | - | 7 | Usability Analyst |
| **E7** | USA | MSc in HCI | - | - | 5 | Interaction Designer |
| **E8** | Spain | MSc in HCI | - | - | 5 | UX/Usability Designer |

**Table 2. Profiles of Novice Crowd Usability Inspectors**

| **Experiment I – WHO-EU Web Dashboard** | | | | | | |
|---|---|---|---|---|---|---|
| **Resource ID** | **Country** | **English Language Skills** | **Qualification** | **Experience in Software Testing (Years)** | **Familiarity with HCI/Usability Experience** | **Familiarity with WHO-EU Web Dashboard (before Inspection)** |
| **C1** | Ukraine | Non-Native | Diploma in Computer Science | 2 | performed a Usability Assignment | Nil |
| **C2** | Armenia | Non-Native | Bachelors in IT | 1 | Familiar with Quality Assurance Testing | Nil |
| **C3** | Kenya | Non-Native | BSc. Computer Science | Nil | Studied HCI as Course/ No Experience | Nil |
| **C4** | India | Non-Native | BSc. Computer Applications | 2 | Studied HCI as Course/ No Experience | Nil |
| **C5** | Russia | Non-Native | BSc. Informatics | 2 | Familiar with Quality Assurance Testing | Nil |
| **Experiment II – www.ebay.com & Experiment III – www.booking.com** | | | | | | |
| **Resource ID** | **Country** | **English Language Skills** | **Qualification** | **Experience in Software Testing (Years)** | **Familiarity with HCI/Usability** | **Familiarity with Test Objects (before Inspection)** |
| **C6** | Turkey | Non-Native | Bachelor of Arts | 2 | Attended an HCI Course/ 2 years' Usability Testing Experience | Yes |
| **C7** | Pakistan | Non-Native | Bachelors in IT | 1.5 | Familiar with Quality Assurance Testing | Familiar with similar websites |
| **C8** | UK | Native | Bachelor of Business Administration | 2 | Nil | Not frequent user of www.eBay.com /No experience with www.booking.com |
| **C9** | Russia | Non-Native | BSc. Computer Science | 2 | Studied HCI as Course/ No Experience | Familiar with similar websites |
| **C10** | India | Non-Native | BSc. Computer Science | 1 | Studied HCI as Course/ No Experience | Familiar with similar websites |
| **C11** | Norway | Non-Native | BSc. Software Engineering | 1 | Studied HCI as Course/ No Experience | Familiar with similar websites |

# 3. ANALYSIS & DISCUSSION

A total of 32 usability inspection reports, including 17 crowd's and 15 experts' evaluations, were submitted to authors for further analysis. It is important to mention that a single usability inspector performed each usability inspection. It took three authors about 130 hours each to analyze a total of 508 comments. We employed Cohen's kappa as an inter-rater agreement measure to assess agreement reliability on coding usability problems (Carletta, 1996). The kappa (k) overall value was observed as 0.87, i.e., almost a perfect agreement. Subsection 3.1 discusses in detail the analysis of usability inspection reports. A total of 209 issues were uniquely reported, out of which 177 issues were reported by expert heuristic usability inspection and 121 issues by novice crowd usability inspection with an overlap of 98 issues (47%). The results contain 32 critical issues and 69 serious issues. An overview of the results is shown in table 3. The overlap ratio of the unique problems between novice crowd inspection and expert heuristic inspection for the first trial of the experiment, i.e., 54% (46 out of 85 unique issues), was higher compared to the second trial overlap ratio, i.e., 39% (25 out 64 unique issues) and third trial overlap ratio, i.e., 45% (27 out of 60 unique issues) of the experiment. One possible explanation for this difference is that the test object for the first trial of the experiment was a web dashboard containing a single-screen interface leading to a higher rate of finding similar usability issues. On the other hand, the test objects for the second and third trials of the experiment were two transactional websites, i.e., www.ebay.com, and www.booking.com, respectively, having multiple web pages with different functionalities, leading to lower chances of finding similar usability issues.

**Table 3. Overview of results**

| | WHO-EU Web Dashboard | www.ebay.com | www.booking.com |
|---|---|---|---|
| Original comments | 187 | 160 | 161 |
| Comments after splitting & combining | 193 | 163 | 169 |
| Total number of unique issues | 85 | 64 | 60 |
| Minor issues | 24 | 46 | 38 |
| Serious issues | 44 | 9 | 16 |
| Critical issues | 17 | 9 | 6 |
| Experts identified issues | 76 | 53 | 48 |
| Crowd identified issues | 55 | 33 | 33 |
| Issues overlapped (Crowd+Expert) | 46 | 25 | 27 |

### 3.1 Analyzing Usability Inspection Reports

We adopted a systematic approach to analyzing the individual usability inspection reports to code usability problems. Figure 2 shows a stepwise process for analyzing usability inspection reports.

**Step 1 – Identifying Problem Comments:**

The authors analyzed usability inspection reports for problem comments. In this regard, the authors encoded problem comments within the usability problem description. However, complex usability problem descriptions can have multiple unique problem comments and need further processing (Molich and Dumas, 2008).

**Step 2 – Splitting and Combining Problem Comments:**

The composite problem comments were split into multiple atomic problem comments (Molich and Dumas, 2008). They were split manually, extracting sentences topic-wise until they could not be further divided. Xu et al. (2015) also used the topic modeling technique to categorize crowd feedback. Moreover, related but non-identical problem comments were combined to represent a single problem comment. In this way, the usability inspection report becomes more understandable and manageable. However, each atomic comment was equal to the original comment reported in the usability inspection report. Moreover, each atomic problem comment has its solution to fix it without affecting any other atomic problem comments (Molich and Dumas, 2008). The following example demonstrates how a composite original problem comment was split into multiple atomic problem comments from the 1st trial of the experiment, i.e., WHO-EU Web Dashboard:

*The menu that appears after clicking "Select data" and the table with core health indicators list the same indicators. Selecting an option on the "Select data" menu does not affect the health indicators' table. It is not self-evident which way of selecting a health indicator a user should choose and how to start.*

1) *The menu that appears after clicking "Select data" and the table with core health indicators list the same indicators.*
2) *Selecting an option on the "Select data" menu does not affect the health indicators' table.*
3) *It is not self-evident which way of selecting a health indicator a user should choose and how to start.*

A total of 34 original comments out of 508 were split into multiple atomic comments. In 46 cases, we combined multiple atomic comments in the same inspection report as found them identical. A total of 27 inspectors reported identical comments.

**Step 3 – Identifying & filtering false positives:**

A false alarm is an incorrectly identified usability problem (Gray and Salzman, 1998). In other words, a false alarm is a usability problem solving that would not improve the system's usability (Molich and Dumas, 2008). They further argued that if the myth that expert reviews find many false alarms were valid, there would have been an unevenly large number of usability problems reported by expert reviews than user-based methods (Molich and Dumas, 2008). Some authors believe that the usability issues reported by expert evaluations not confirmed by user-based methods are false alarms. This argument would have been reasonable if the results of user-based methods were reproducible (Molich and Dumas, 2008).

Molich and Dumas (2008) believe that the main reason for earlier false alarms in expert reviews was that too few user-based tests were conducted to verify the usability problems reported by expert reviews. Most of the time, there is only one usability test involving an evaluator of insufficient experience with few users and tasks.

However, three authors reviewed all the usability issues identified by usability inspectors for false positives, based on the definition provided in subsection 2.2.3 Instrumentation. A total of 58 false positives were identified, out of which eight were found in expert heuristic usability inspection while 50 were found in novice crowd usability inspection (see Table 4). False positives found in novice crowd usability inspection were substantially more than the expert heuristic usability inspection. The apparent reason for the substantial difference between the two UEMs regarding false positives is usability experts' profiles and novice crowd usability inspectors, i.e., skills, experience, and education.

**Step 4 – Combining similar usability problems at a higher abstraction level:**

Usability problems may be defined at different levels of abstraction by the same evaluator and multiple evaluators simultaneously in their inspection reports (Howarth et al., 2009). Combining similar usability problems existing at different abstraction levels is necessary to avoid over and under-reporting usability problems (Cockton and Lavery, 1999). For example, suppose one usability problem comment may report that a website is cluttered with overwhelming information. Another problem comment may describe that the property information page is cluttered with an over-dose of irrelevant information. Therefore, the master list should contain usability problems representing a higher abstraction level. In comparison, individual inspection reports containing lower abstraction levels may be mapped with a master list to describe the same usability problem.

**Step 5 – Filtering identical usability problems from Master-List:**

Finally, identical problem comments reported by multiple usability inspectors were reported uniquely in the master list. Two or more problem comments were considered identical if fixing the problem in one comment will fix the problem in other comments (Molich & Dumas, 2008).

Few, sample identical problem comments from different usability inspectors, reported in www.booking.com, are listed below:

***www.booking.com***:

1. *Searching Accommodation: Within each property block, there are several rating systems. Which does a user follow? What does each mean? are they all related? It is quite confusing to understand the rating system.*
2. *Searching Accommodation: Different schemes of rating violate best practices of usability.*
3. *Searching Accommodation: Stars' rating option is confusing.*
4. *Searching Accommodation: There is a usability consideration over the meaning of star ratings. There is no easy way to get info on star ratings from the property page.*

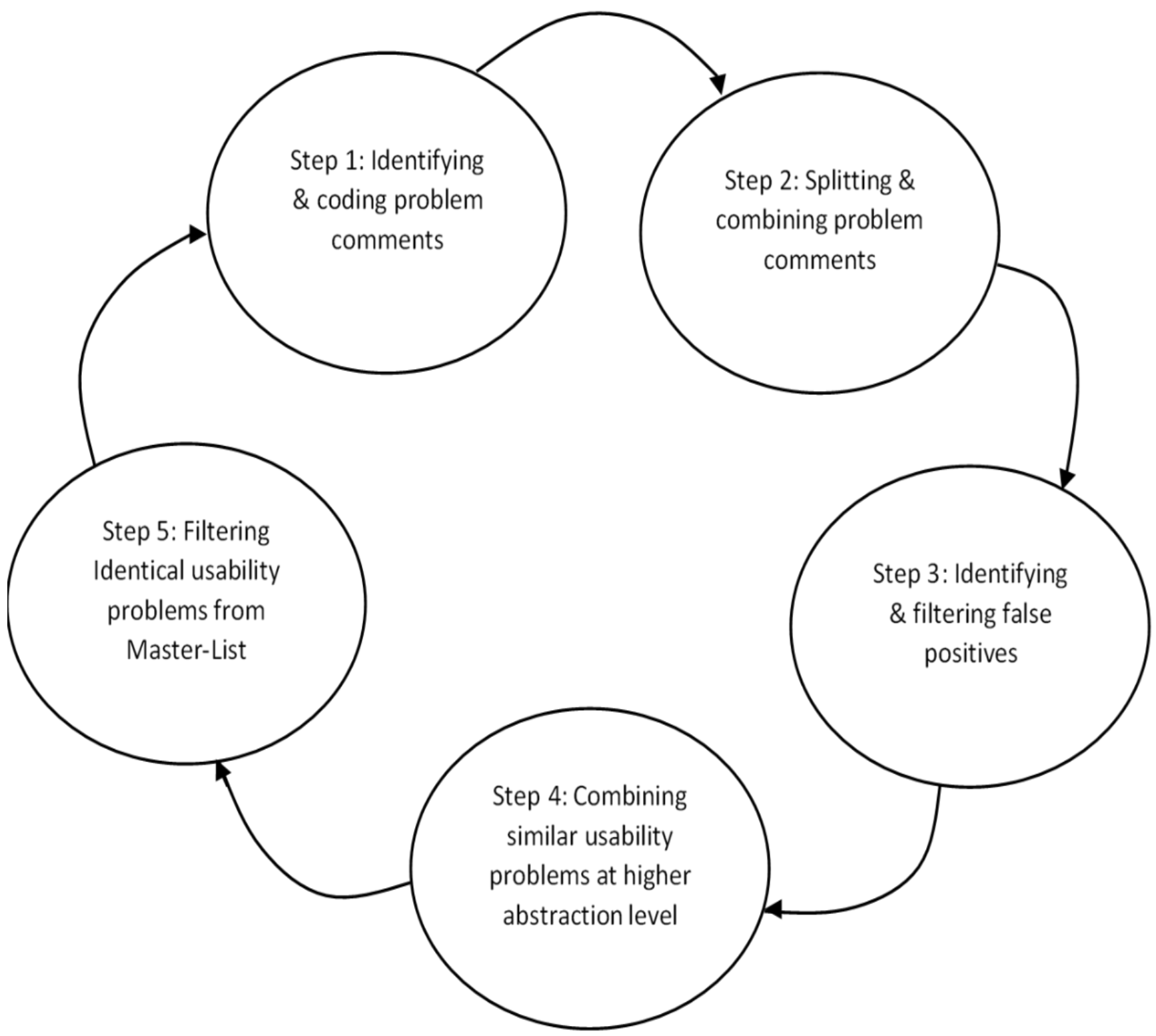

**Figure 2: Analyzing usability inspection reports.**

## 3.2 Severity of Usability Problems

We used expert heuristic usability inspection as a benchmark for comparison. The severity of novice crowd inspector's reported issues was derived from experts' reported problems that matched with them, except for 23 crowd usability issues that did not match. Therefore, no contradiction was observed for the severity of the usability issues between novice crowd usability inspection and expert heuristic usability inspection.

In expert heuristic usability inspection, 51 issues were reported by multiple experts. For 34 out of 51 issues, the majority of the experts suggested the classification as serious, while deviations were only one degree away from the average classification, which we considered acceptable. For the rest of the usability issues, classification varied from minor to critical, and in this case, averaged severity was considered. No other contradictory classifications were observed.

### 3.3 Key Issues

Molich and Dumas (2008) suggested that if multiple experts categorize an issue as serious or critical, it indicates that such an issue may have considerable design consequences. Hence, we considered such issues as a quality measure based on the expert's judgment. Moreover, we considered a key usability issue as a repeatedly identified serious usability issue. In other words, a key usability issue is: 1) is at least reported by three usability experts or at least two usability experts and one or more crowd inspectors. 2) is identified as a serious or critical issue on average. We calculated average classification by associating values 1,2,3 to minor, serious, and critical issues, respectively. We then took an average of the values and rounded it off. A total of 30 key usability issues were reported uniquely. Details of the key usability issues reported by each inspector are shown in table 4. Some examples of the key usability issues are as below:

***www.booking.com:***

1. *The property detail page looks cluttered with too much information and a poorly organized content hierarchy.*
2. *There are different rating symbols displayed when you view any accommodation, i.e., stars, thumbs-up, squares. These various rating schemes are not consistent and are quite confusing and difficult to comprehend.*

***www.ebay.com:***

1. *The user is not able to select multiple items within one refiner category (e.g., select AT&T and T-Mobile under Network). It is a UX glitch that hinders the refiner experience. Users are immediately led to a page with the results before they can select any other preferences.*
2. *The page has focused on money/payment, not the product. I want to know if I want the product before I start to think about the price. They should place a description together with the product, not far below. There is a risk of missing the description and not bidding on the product of my interest.*

A list of all the key usability issues is provided in Appendix D. Most of the key usability issues were identified in the WHO-EU web dashboard 70% (21 out of 30). In comparison, 17% (5 out of 30) were found on booking.com, and the rest of the 13% (4 out of 30) were reported on eBay.com. The key usability issue identified by the maximum number of inspectors in the WHO-EU web dashboard, i.e., 7, including three experts and four novice crowd inspectors, was #12, violating heuristic, Minimize occlusion. Likewise, key issue#2 received maximum hits (8 inspectors, including four experts and four novice crowd inspectors) in Booking.com, violating heuristic, Aesthetic and minimalist design. For ebay.com, the most identified key issue was #3 (identified by seven inspectors, including four experts and three novice crowd inspectors), violating heuristic, Consistency, and standards.

**Table 4. Detailed results**

**Experiment I – WHO-EU Web Dashboard**

| Resource | E1 | E2 | E3 | E4 | E5 | C1 | C2 | C3 | C4 | C5 |
|---|---|---|---|---|---|---|---|---|---|---|
| Total number of original comments | 26 | 22 | 17 | 17 | 22 | 17 | 19 | 15 | 18 | 14 |
| Total number of comments after splitting & combining | 29 | 24 | 15 | 18 | 21 | 15 | 15 | 19 | 22 | 15 |
| Total number of false positives | 0 | 1 | 0 | 0 | 1 | 3 | 1 | 2 | 3 | 2 |
| Total number of usability problems | 29 | 23 | 15 | 18 | 20 | 12 | 14 | 17 | 19 | 13 |
| Total number of minor problems | 3 | 3 | 2 | 2 | 2 | 3 | 5 | 3 | 2 | 5 |
| Total number of serious problems | 19 | 15 | 10 | 13 | 10 | 4 | 3 | 11 | 13 | 4 |
| Total number of critical problems | 7 | 5 | 3 | 3 | 8 | 5 | 6 | 3 | 4 | 4 |
| Total number of key usability issues | 13 | 7 | 12 | 12 | 12 | 7 | 8 | 5 | 10 | 3 |
| Person hours used | 13 | 17 | 12 | 14 | 16 | 5.6 | 4.1 | 6.1 | 4.1 | 5.6 |
| Payment made to each inspector (USD) | 300 | 400 | 200 | 250 | 350 | 5 | 5 | 6 | 7 | 5 |
| Total cost per inspection (USD) | 300 | 400 | 200 | 250 | 350 | 65 | 65 | 66 | 67 | 65 |
| Cost per issue (USD) | 10.3 | 17.4 | 13.3 | 13.9 | 17.5 | 5.4 | 4.6 | 3.9 | 3.5 | 5.0 |
| Cost per key issue (USD) | 23.1 | 57.1 | 16.7 | 20.8 | 29.2 | 9.3 | 8.1 | 13.2 | 6.7 | 21.7 |
| Issues per hour | 2.2 | 1.4 | 1.3 | 1.3 | 1.3 | 2.1 | 3.4 | 2.8 | 4.6 | 2.3 |
| Key issues per hour | 1.0 | 0.4 | 1.0 | 0.9 | 0.8 | 1.3 | 2.0 | 0.8 | 2.4 | 0.5 |
| Validity of usability problems | 100% | 96% | 100% | 100% | 95% | 80% | 93% | 89% | 86% | 87% |
| Thoroughness of usability problems | 34% | 27% | 18% | 21% | 24% | 14% | 16% | 20% | 22% | 15% |

**Experiment II – www.ebay.com**

| Resource | E1 | E2 | E6 | E7 | E8 | C6 | C7 | C8 | C9 | C10 | C11 |
|---|---|---|---|---|---|---|---|---|---|---|---|
| Total number of original comments | 20 | 17 | 15 | 13 | 8 | 17 | 15 | 18 | 14 | 16 | 07 |
| Total number of comments after splitting & combining | 21 | 16 | 14 | 11 | 8 | 19 | 20 | 22 | 14 | 18 | 0 |
| Total number of false positives | 0 | 1 | 1 | 0 | 2 | 5 | 3 | 4 | 3 | 3 | 4 |
| Total number of usability problems | 21 | 15 | 13 | 11 | 6 | 14 | 17 | 18 | 11 | 15 | 3 |
| Total number of minor problems | 7 | 11 | 12 | 6 | 3 | 5 | 4 | 11 | 3 | 6 | 0 |
| Total number of serious problems | 7 | 3 | 0 | 3 | 3 | 5 | 6 | 3 | 4 | 3 | 0 |
| Total number of critical problems | 7 | 1 | 1 | 2 | 0 | 4 | 7 | 4 | 4 | 6 | 0 |
| Total number of key usability issues | 3 | 2 | 1 | 1 | 3 | 1 | 3 | 3 | 2 | 3 | 1 |
| Person hours used | 8 | 5 | 5 | 7 | 8 | 3 | 2.6 | 7.6 | 3 | 2.8 | 1 |
| Payment made to each inspector (USD) | 300 | 400 | 250 | 200 | 200 | 5 | 6 | 7 | 5 | 6 | 1 |
| Total cost for per inspection (USD) | 300 | 400 | 250 | 200 | 200 | 65 | 66 | 67 | 65 | 66 | 61 |
| Cost per issue (USD) | 14.3 | 26.7 | 19.2 | 18.2 | 33.3 | 4.6 | 3.9 | 3.7 | 5.9 | 4.4 | 20.3 |
| Cost per key issue (USD) | 100 | 200 | 250 | 200 | 66.7 | 65 | 22 | 22.3 | 32.5 | 22 | 61 |
| Issues per hour | 2.6 | 3.0 | 2.6 | 1.6 | 0.8 | 4.7 | 6.5 | 2.4 | 3.7 | 5.4 | 3 |
| Key issues per hour | 0.4 | 0.4 | 0.2 | 0.1 | 0.4 | 0.3 | 1.2 | 0.4 | 0.7 | 1.1 | 1 |
| Validity of usability problems | 100% | 94% | 93% | 100% | 75% | 74% | 85% | 82% | 79% | 83% | 43% |
| Thoroughness of usability problems | 34% | 25% | 21% | 18% | 10% | 23% | 28% | 30% | 18% | 25% | 5% |

**Experiment III – www.booking.com**

| Resource | E1 | E2 | E6 | E7 | E8 | C6 | C7 | C8 | C9 | C10 | C11 |
|---|---|---|---|---|---|---|---|---|---|---|---|
| Total number of original comments | 15 | 12 | 14 | 12 | 15 | 17 | 15 | 16 | 16 | 21 | 8 |
| Total number of comments after splitting & combining | 17 | 13 | 13 | 11 | 16 | 19 | 13 | 20 | 16 | 24 | 7 |
| Total number of false positives | 0 | 0 | 1 | 1 | 0 | 4 | 3 | 3 | 3 | 3 | 1 |
| Total number of usability problems | 17 | 13 | 12 | 10 | 16 | 15 | 10 | 17 | 13 | 21 | 6 |
| Total number of minor problems | 6 | 10 | 8 | 8 | 12 | 4 | 4 | 5 | 8 | 7 | 3 |
| Total number of serious problems | 8 | 1 | 4 | 2 | 4 | 9 | 5 | 9 | 4 | 11 | 2 |
| Total number of critical problems | 3 | 2 | 0 | 0 | 0 | 2 | 1 | 3 | 1 | 3 | 1 |
| Total number of key usability issues | 3 | 2 | 4 | 2 | 3 | 5 | 2 | 3 | 2 | 5 | 3 |
| Person hours used | 7 | 6 | 5 | 6 | 8 | 2.5 | 3 | 5.6 | 2.6 | 2.6 | 1 |
| Payment made to each inspector (USD) | 300 | 400 | 250 | 200 | 200 | 7 | 5 | 6 | 5 | 8 | 3 |
| Total cost per inspection (USD) | 300 | 400 | 250 | 200 | 200 | 67 | 65 | 66 | 65 | 68 | 63 |
| Cost per issue (USD) | 17.6 | 30.8 | 20.8 | 20.0 | 12.5 | 4.5 | 6.5 | 3.9 | 5.0 | 3.2 | 10.5 |
| Cost per key issue (USD) | 100 | 200 | 62.5 | 100 | 66.7 | 13.4 | 32.5 | 22 | 32.5 | 13.6 | 21 |
| Issues per hour | 2.4 | 2.2 | 2.4 | 1.7 | 2.0 | 6.0 | 3.3 | 3 | 5 | 8.1 | 6 |
| Key issues per hour | 0.4 | 0.3 | 0.8 | 0.3 | 0.4 | 2.0 | 0.7 | 0.5 | 0.8 | 1.9 | 3 |
| Validity of usability problems | 100% | 100% | 92% | 91% | 100% | 79% | 77% | 85% | 81% | 88% | 75% |
| Thoroughness of usability problems | 31% | 24% | 22% | 19% | 30% | 28% | 19% | 31% | 24% | 39% | 11% |

### 3.4 Resource Usage

How efficiently resources are used is an essential aspect of comparing UEMs. If we look at Figure 3, we can see that experts have outnumbered novice crowd inspectors in terms of identifying key issues per person. However, person-hours used to identify key usability issues are not free. The person-hours used to change the whole view altogether. The key usability issues identified per hour by each resource are shown in Figure 3. We can see that novice crowd inspectors are more efficient than experts in identifying key usability issues per hour. Figure 4 indicates that experts have identified more key usability issues at the cost of more time. Likewise, Figure 4 shows that experts' key usability issues were quite expensive compared to crowd inspectors. It is worth mentioning that novice crowd inspectors identified 27 out of the total 30 key usability issues.

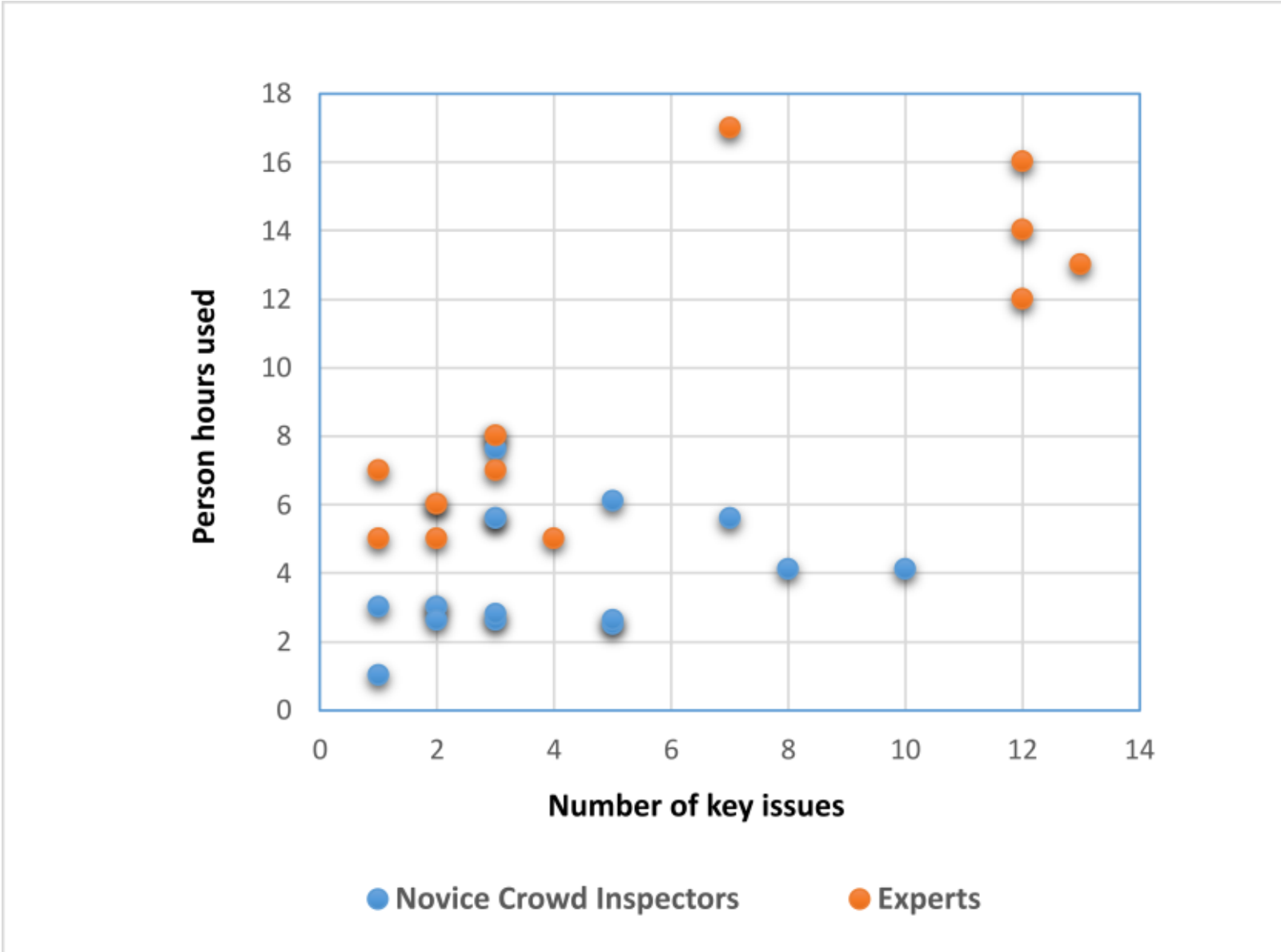

Figure 3 Crowd vs. Experts – Person hours used.

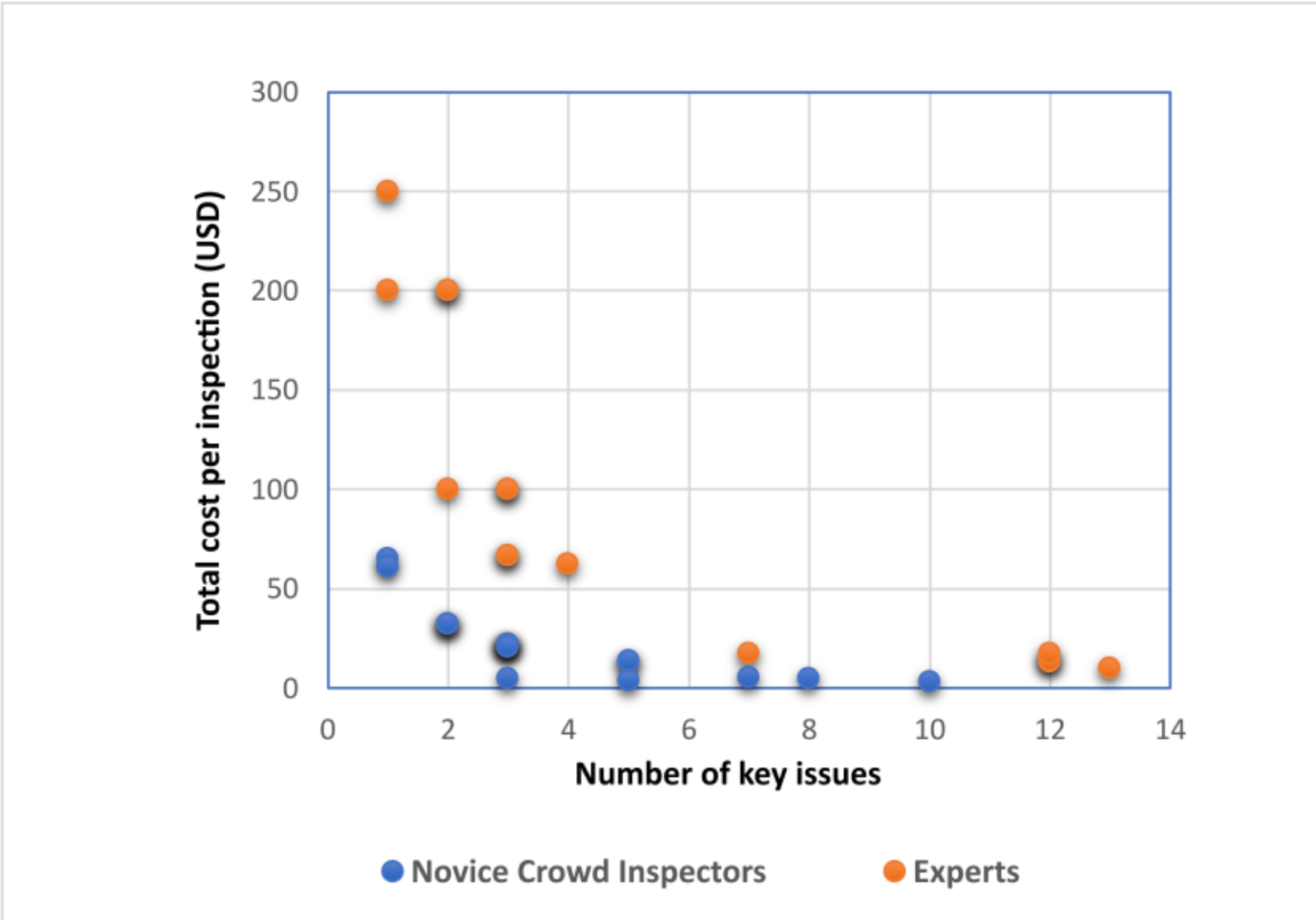

Figure 4: Crowd vs. Experts – Cost per inspection

### 3.5 Statistical Analysis

To test the hypotheses (formulated above in subsection 2.2 Planning) statistically, we performed significance testing. The results for each hypothesis are discussed below:

**$H_a$: Usability Issues Found**

To find out whether novice crowd usability inspection finds the same usability issues (w.r.t. content & quantity) or different, we need to split this hypothesis into two parts, i.e., a) content b) quantity. As far as the content is concerned, we have already discussed that 47% (98 out of 209) of total unique issues were the same and identified by both novice crowd usability inspection and expert heuristic usability inspection. Moreover, in terms of key usability issues, 90% (27 out of 30) of total key issues overlapped between novice crowd usability inspection and expert heuristic usability inspection. To test the second part of the hypothesis, i.e., quantity, we compared the two samples for the number of usability issues identified, using a non-parametric test, the Mann-Whitney U test. The result ($p=0.83$) supports the alternate hypothesis $H_{a1}$, and there were no significant differences between the two groups. In other words, we can say that novice crowd usability inspection, on average, finds the same usability issues in terms of content & quantity as expert heuristic usability inspection.

**$H_b$: Cost-Effectiveness**

It is worth mentioning that the cost for novice crowd inspectors is aggregate; see table 4. And it includes the cost for expert heuristic usability inspection conducted by a single usability expert (E1). We divided this cost on all novice crowd inspectors, adding 60 USD for each crowd inspector in each experiment's trial. We calculated the cost for novice crowd usability inspection in this way since almost half of the use-cases were based on the problems identified by expert E1 in each trial.

To compare the UEMs for cost-effectiveness, we considered cost per issue identified for both samples, using the Mann-Whitney U test. The result, i.e., $p=0.00001$, supports the alternate hypothesis that novice crowd usability inspection is more economical than expert heuristic usability inspection is true.

**$H_c$: Time-Efficiency**

Our third hypothesis was based on the time spent on a usability inspection process. The time spent on the usability inspection process for each evaluation is shown in table 4. The time duration for novice crowd usability inspection is aggregate. It includes a) the time for novice crowd inspection, b) the time spent on expert heuristic usability inspection conducted by a single expert (E1), and c) the time for designing use-cases for crowd usability inspection (i.e., 3 hours). The time duration for prerequisite expert review and use-cases' designing was divided on all novice crowd usability inspections for each experiment's trial.

The job completion time was considered by comparing the two samples. The result ($p=0.00001$) reveals that there is a significant difference between the two samples. Hence, the alternate hypothesis, i.e., $H_{c1}$ – Novice crowd usability inspection on average takes less time than expert heuristic usability inspection, is true. A summary of the hypothesis testing results is shown in table 5.

Table 5. Hypothesis Testing Results

| S.No. | Hypothesis | P-value | Significance | Result |
|---|---|---|---|---|
| 1 | $H_a$: Usability issues found | p=0.83 | Not significant | $H_{a1}$ - Usability inspection using crowdsourcing on average finds the same usability issues (w.r.t. content & quantity) as a heuristic usability inspection |
| 2 | $H_b$: Cost-Effectiveness | p=0.00001 | Significant | $H_{b1}$ - Usability inspection using crowdsourcing incurs less cost than heuristic usability inspection. |
| 3 | $H_c$: Time-Efficiency | p=0.00001 | Significant | $H_{c1}$ - Usability inspection using crowdsourcing, on average, takes less time than heuristic usability inspection. |

## 3.6 Comparison of Results

Usability inspection employing novices has been investigated by researchers in the past (Koutsabasis et al., 2007; Følstad et al., 2010). The performance of the evaluators was often compared using two measures, i.e., validity and thoroughness. The validity of the results can be obtained by dividing the total number of correct problem predictions (real problems) by the total number of problem predictions, including false positives (All comments). The validity of the results shows the number of correct problem predictions, excluding false positives (Følstad et al., 2010). The thoroughness of the results can be measured by dividing the total number of correct problem predictions using all the real problems. The thoroughness of the results shows the degree to which real problems have been identified compared to all problems (Følstad et al., 2010).

The novice crowd usability inspection results are encouraging in terms of the validity of the results compared with expert heuristic usability inspection and other studies (Koutsabasis et al., 2007, Følstad et al., 2010) see table 4, 6. The validity of the novice crowd usability inspection results remained relatively consistent in all three experiments, ranging from 82% to 87%, with an average of 83% if we exclude the work of the crowd usability inspector, C11 (outlier). However, after including the work of C11, validity drops to a level ranging from 74% to 87%, with an average of 81%. In terms of thoroughness, novice usability inspection results are promising compared to expert heuristic usability inspection (see table 6). Compared with other studies, novice inspections are encouraging compared with Koutsabasis et al. (2007), while less thorough than the results of Følstad et al. (2010). One reason for the high level of thoroughness in the results of Følstad et al. (2010) could be that their team of work domain experts comprised three times more participants, i.e., 15, compared to our study, i.e., 5-6. The validity and thoroughness of crowd inspectors and experts' usability problems are also provided in table 4.

## 3.7 Challenges of Crowdsourcing and Solution Strategies.

Over time, researchers have reported several challenges faced by crowdsourcing, including low quality of work (Zhao & Zhu, 2014, Wang et al., 2017), the cognitive load of examining the quality of work (Zhao & Zhu, 2014), the validity of self-reported demographic details (Liu et al., 2012), malicious (Wang et al., 2017) and anonymous submission of work multiple times to increase the reward, lack of appropriate reward (Guaiani, and Muccini, 2015, Garcia-Molina et al., 2016) and review mechanism of work.

Crowdsourcing has matured quite a bit in the field of computing (Ambreen and Ikram, 2016). Modern crowdsourcing platforms, i.e., Upwork, are quite established and offer different ways and means to cope with the challenges discussed earlier. We will discuss these features and support modern crowdsourcing platforms learned in this study while taking software usability evaluation into context.

**Table 6. Comparison of results**

| Study Reference | Usability Evaluation Method | Teams Size | Validity | Thoroughness |
|---|---|---|---|---|
| (Koutsabasis et al., 2007) | Heuristic Evaluation | 3 | 94.4% | 24.3% |
| | Heuristic Evaluation | 3 | 60.7% | 24.3% |
| | Heuristic Evaluation | 3 | 100% | 20% |
| | Cognitive Walkthrough | 3 | 85.7% | 25.7% |
| | Cognitive Walkthrough | 3 | 70.8% | 24.3% |
| | Think-aloud Protocol | 3 | 90.5% | 27.1% |
| | Think-aloud Protocol | 3 | 94.4% | 24.3% |
| | Think-aloud Protocol | 3 | 82.4% | 20% |
| | Co-Discovery | 3 | 74.4% | 41.4% |
| (Følstad et al., 2010) | Work Domain Experts | 15 | 22.2% | 100% |
| | Work Domain Experts | 15 | 22% | 50% |
| | Work Domain Experts | 15 | 32% | 53% |
| | Work Domain Experts | 15 | 56% | 83% |
| This study | Crowd Inspection (Web Dashboard) | 5 | 87% | 18% |
| | Crowd Inspection (eBay.com) | 6 | 74% | 22% |
| | Crowd Inspection (Booking.com) | 6 | 81% | 25% |
| | Expert Inspection (Web Dashboard) | 5 | 98% | 25% |
| | Expert Inspection (eBay.com) | 5 | 92% | 22% |
| | Expert Inspection (Booking.com) | 5 | 97% | 25% |

**The validity of self-reported demographics.** Currently, crowdsourcing platforms maintain verified profiles of crowd-workers, e.g., Upwork issues identity badges for verified accounts (Upwork, 2020). Earning an identity badge requires two-stage verification is performed, a) video call verification, b) verification through government-issued ID. Moreover, maintaining crowd workers' profiles shows how they performed in previous projects and their feedback and ratings from employers (Upwork, 2020). Zhao and Zhu (2014) also suggested a public rating-based strategy to control the quality of the work. Besides, providing online tests to crowd-workers to prove their skills is another feature of crowdsourcing platforms (Upwork, uTest, 2020).

**Malicious and anonymous work.** Although some crowdsourcing platforms allow crowd-workers to submit their work anonymously (MTurk, 2020), the same crowd-worker may complete the same task multiple times to increase their reward. However, other crowdsourcing platforms (e.g., Upwork) are very strict about crowd workers' identity, and it is almost impossible to complete the same task multiple times to increase the reward.

**Poor quality of work.** Assessing crowd workers' quality is a major challenge in crowdsourcing (Zhao & Zhu, 2014). Especially, usability problem descriptions reported by novices need to be examined by usability experts for false positives (de Lima Salgado et al., 2018). Moreover, novice usability inspectors need to be supported with use-cases, etc. (de Lima Salgado et al., 2018). In our study, we have used a similar approach and supported novice crowd usability inspectors with use-cases. Besides, we compared the usability problem descriptions reported by novice crowd usability inspectors with experts to determine the severity of usability problems. Furthermore, the authors' panel examined the usability problem descriptions for false positives and duplicate comments (Molich et al., 2004).

**Lack of review and reward mechanism** Some crowdsourcing platforms (e.g., MTurk) do not allow to review of the work submitted by crowd workers, and crowd workers get paid for that work, regardless of the quality of work. However, this is not the case with all the crowdsourcing platforms. Some crowdsourcing platforms (e.g., Upwork) provide a proper mechanism to review the work submitted by crowd workers and ask them to revise the work submitted if required. And the payment is released only when the employer is satisfied with the quality of the work, and the employer has an option to discard the work altogether with no payments made to crowd workers.

Moreover, employers can hire crowd workers on a fixed-price contract or per hour rate. Employers can divide the work into milestones and can release payments with milestones after reviewing the work. Hence, reward and review mechanisms are well controlled in modern crowdsourcing platforms (e.g., Upwork).

### 3.8 Wisdom of crowd and novice usability inspection

Current literature on usability/interactive design evaluation employing crowdsourcing has several limitations to the best of our knowledge. The existing literature does not address the challenges of crowdsourcing nor proposes any solution strategies in this regard, except comparing the performance of crowd-workers with the control group (see subsection 1.2.3 and 1.2.4 for more details). Finally, subsection 3.7 presents the experiences learned from this study to overcome crowdsourcing challenges while employing novice crowd usability inspectors for usability evaluation.

Moreover, the findings of our study imply that crowd-workers with HCI/interactive design skills and assistive tools can produce results comparable to experts. For example, we employed uses-case as assistive tools while Yuan et al. (2016) used a rubric of design principles. However, the study by Yuan et al. (2016) was based on a design critique of UI elements compared to our study based on comparative usability inspection. On the other hand, the studies that did not support crowd-workers with assistive tools and HCI/interactive design skills could not produce results comparable to experts (Xu et al., 2015, Bruun & Stage, 2015).

Besides, the studies conducted without employing crowdsourcing also support the argument that novices without experience and knowledge of usability would not produce comparable results to usability experts (Nielsen,1992; Borys and Laskowski, 2014). On the other hand, the studies that equipped experienced novices with assistive tools produced encouraging results (Howarth et al., 2009; Følstad et al., 2010). Some of the studies on novice usability evaluation did not practically validate their work. However, these studies also supported the argument that novices should be supported with assistive tools, i.e., use-cases, storyboards, design patterns, etc. (Botella et al., 2013; De et al., 2018).

Moreover, our study contributes to existing knowledge on crowd usability evaluation. We have not found any other study comparing novice usability inspectors (equipped with use-cases and HCI skills) with usability experts using crowdsourcing (see subsection 1.2.5 for more on the novelty of this study).

In the context of the above discussion and results of our study, it can be safely affirmed that the crowd's wisdom has the potential for novice usability inspection. Moreover, experienced novices/beginners with assistive tools can perform comparably to usability experts. Although novice crowd usability inspection gives comparable results to expert heuristic usability inspection, it is a hybrid method that involves a usability expert. We do not claim that crowd usability inspection can always be the best substitute for usability testing or expert usability inspection. It can be adopted as an economical alternate solution for budget-constrained software development organizations for usability evaluation. When it is expensive to hire 3-5 usability experts or conduct usability testing, novice crowd usability inspection may be employed as an alternative method for usability evaluation. Unlike expert usability inspection, novice crowd usability inspection involves the following overheads:

a) Designing use-cases to develop a questionnaire for novice crowd usability inspection.
b) Analyzing usability comments to extract usability problems, filtering false alarms, and determining the usability problems' severity.

In comparison to expert usability inspection, novice usability evaluators need to be supported with use-cases. However, designing the use-cases is time-consuming. Nevertheless, tasks are also identified in usability testing as well. Besides, analyzing the usability comments of novice crowd inspectors requires intellectual effort. However, the efforts for analyzing the usability comments can be mitigated using text-mining techniques using machine learning. A similar approach has been suggested by (Xu et al., 2015, Wang et al., 2017).

# 4. LIMITATIONS AND VALIDITY THREATS

A range of validity threats needs to be addressed for a comparative usability study (Gray and Salzman, 1998). We will discuss them one by one.

## 4.1 Statistical Conclusion Validity

There are three main issues in statistical conclusion validity: low statistical power and wildcard effect, random heterogeneity in sample, and too many comparisons to capitalize on chance factors. For expert evaluation, 3-5 experts are recommended (Jeffries and Deservers, 1992), and we employed five usability experts to perform a heuristic expert evaluation for each trial of the experiment. We did not notice a wildcard effect in expert heuristic evaluation results due to the random heterogeneity of experts' profiles. Therefore, we did not discard any results for expert heuristic evaluation. For novice crowd usability inspection, we analyzed the results for any outliers as adopted in other comparative usability studies (Chattratichart and Lindgaard, 2008; Gomide et al., 2014). The crowd usability inspector C11 identified significantly fewer usability problems (3 in 1$^{st}$ trial and 6 in 2$^{nd}$ trial of the experiment) than other crowd inspectors. The crowd inspectors, on average, identified 14 usability problems in each trial of the experiment. However, we did not exclude any usability inspection reports to give a more objective view of the results. Furthermore, we did not perform too many comparisons; therefore, there is no possibility of exploiting the chance factor. Besides, we did not rely on eyeball tests, i.e., averages, percentages; instead, we performed statistical tests to prove the validity of hypotheses.

## 4.2 Internal Validity

Internal validity threats concern whether the differences between the UEMs are causal rather than correlational. There are three main aspects of internal validity, i.e., instrumentation, selection, and setting. There are two main biases in instrumentation, i.e., how the severity of the usability problems is assessed and their categorization. To avoid biases in determining usability problems' severity, each expert independently identified the severity of usability problems. Moreover, the average severity of common usability problems belonging to the same key usability issue was considered regarding the key usability issues. The severity of the usability problems identified by novice crowd inspectors was derived from experts' identified usability problems, which matched with them, leaving only 23 usability problems that did not match. The severity of these 23 usability problems was assessed by the expert (E1), whose usability inspection report was used to design the use-cases for novice crowd usability inspection.

Novice crowd usability inspection consisted of use-cases that were based on a single expert evaluation. However, half of the total use-cases were general use-cases developed based on the test objects' functionality. Besides, novice crowd inspectors were not asked to categorize the usability problems in terms of heuristics or severity. Therefore, novice crowd inspectors identified usability problems irrespective of the expert evaluation, categorizing usability problems for severity and heuristics.

The novice crowd inspectors and experts were self-selected based on pre-defined criteria to overcome the selection bias. All the inspectors were hired using the same platform to avoid the problems with settings, i.e., Upwork, and the same platform was used to communicate with them.

## 4.3 Construct Validity

Construct validity deals with two main issues: a). Are we manipulating what we have claimed to manipulate (Causal construct validity)? b). Do we measure what we claimed to measure (effect construct validity)?

In causal construct validity, there are many threats. The most prominent threat is defining, operationalization, and understanding the UEM. Other threats include applying a flexible usability evaluation in only one way (mono-operation bias), applying a UEM to only one type of software (mono-method bias), and learning effect due to treatments' interaction.

To overcome the misinterpretation in the definition of UEMs, we clearly defined both expert heuristic usability inspection and novice crowd usability inspection. The expert evaluation was based on heuristics, while crowd evaluation was based on use-cases with clear instructions, how to perform it. Simple and clear instructions were given to evaluators to avoid mono-operation bias and variations in UEMs' operation. For example, all the experts were asked to perform evaluations independently without interacting with each other. To address the mono-method bias, we have generalized our findings at least to websites and web dashboards. The learning effect was naturalized by avoiding the interaction of the treatments. None of the participants was given both treatments in the experiment, i.e., expert heuristic usability inspection and novice crowd usability inspection.

In effect-construct validity, problems arise when results of empirical UEMs are compared with analytic UEMs. However, this is not the case in this study, and we compared intrinsic features and pay-off measures of expert heuristic usability inspection with novice crowd usability inspection separately. Moreover, we compared the UEMs based on the contents of usability problems identified rather than just the number of problems identified.

## 4.4 External Validity

This experiment was conducted using experts that had relevant knowledge and practical experience of heuristic usability inspection. Besides, novice crowd inspectors were also hired using a crowdsourcing platform, i.e., Upwork. Moreover, real-life systems, i.e., WHO-EU Web

Dashboard and websites like www.booking.com and www.ebay.com, were employed for usability evaluations. Therefore, this research study's findings are valid for novice crowd usability inspection and expert heuristic usability inspection.

### 4.5 Conclusion validity

Our claims are consistent with the hypotheses that have been investigated in this research study. All the claims are compatible with the results of the study.

### 4.5 Limitations of the study

This study has the following limitations due to usability practitioner's effect:

1. The authors designed the questionnaires for novice crowd usability inspection based on a single expert evaluation. Hence, it may introduce limitations such as (a) The expert's usability inspection report may influence the results of crowd usability inspection, (b) The HCI knowledge, experience, skills of usability practitioners may affect the design of the questionnaire for novice crowd usability inspection, and hence results may vary.
2. The authors analyzed crowd usability inspection reports to extract usability problems. There might be a variation in results when a usability practitioner would extract usability problems from novice inspection reports.
3. Usability experts were not asked to have a consensus meeting. Hence, the lack of a consensus meeting might have influenced the results. Jakob Nielsen in 1992 found that double experts can perform better than regular experts. A consensus meeting among usability experts could improve the results, i.e., finding more key issues. However, further research work is required to investigate the effect of consensus meetings among experts.
4. Crowd inspectors were also not asked to have a consensus meeting. A consensus meeting may influence the results. Law and Hvannberg (2008) found that novice usability evaluation in collaborative settings results in inflation and deflation of usability problems.
5. The authors made sure that all the usability inspectors were fluent in the English language. However, we did not investigate the effect of nationality and cultural issues in usability evaluation in this study. Nationality, English language skills, and cultural issues may also influence the results. For example, usability inspectors from countries where English is not a native language may find it difficult to express usability problems. Besides, hiring usability inspectors to evaluate an online shopping website from a country where online shopping is not a cultural trend due to any reason (i.e., poor internet facility, lack of online payment facility, lack of trust in online shopping, etc.) would not be the best choice. Følstad et al. (2010) suggested that domain knowledge can help novice usability inspectors to perform better in usability evaluations.

## 5. CONCLUSION & FUTURE WORK

This research study examined crowdsourcing's potential for novice crowd usability inspection guided by a single expert's heuristic usability inspection and compared it with expert heuristic usability inspection. Expert heuristic usability evaluations and novice crowd usability evaluations were conducted on a crowdsourcing platform, i.e., Upwork. The study involved multiple experiments in generalizing the results, with diverse test objects, including a web dashboard for WHO-EU, and two websites, i.e., www.ebay.com and www.booking.com. Each test object was evaluated by 5-6 novice crowd usability inspectors and five experts to compare the results. Expert usability inspection was based on heuristics, while novice crowd usability inspection was based on a questionnaire (involving use-cases). The authors used a single expert's heuristic usability inspection to design use-cases for novice crowd inspectors. Moreover, the authors analyzed the novice inspection reports to extract usability problems using expert heuristic evaluation as a benchmark. The key findings of this research study are as below:

1. On average, novice crowd usability inspection finds the same usability issues (w.r.t. content & quantity) as expert heuristic usability inspection.
2. Novice crowd usability inspection incurs less cost than expert heuristic usability inspection.
3. Novice crowd usability inspection, on average, takes less time than expert heuristic usability inspection.

Other contributions of the study include:

a) A thorough discussion of challenges in crowdsourcing and their solution was provided in the context of the crowd's wisdom regarding novice usability inspection.
b) The study concludes that equipping novice crowd usability inspectors with use-cases based on a single expert's heuristic usability inspection can help get comparable results from novice crowd usability inspection as expert heuristic usability inspection.
c) The study also developed a method for designing use-cases based on a single expert's heuristic usability evaluation to support novice crowd usability inspection.
d) The study devised a method to analyze novice inspection reports.

This study can be further extended in the future to investigate the following aspects:

a) Actual usability practitioners may be asked to design use-cases for novice crowd usability inspection to investigate the effect of HCI knowledge, experience, and usability practitioners' skills on the quality of use-cases.
b) Novice crowd inspectors and experts may be asked to have a consensus meeting among them in their respective teams to investigate the effect of a consensus meeting on usability inspection.
c) This study can be extended to a sample population of crowd inspectors from different cultures and nationalities to see its impact on usability.
d) Finally, we suggest a need to design a framework for novice crowd usability inspection to guide the practitioners in the industry.

In the context of the findings of this study, we can carefully conclude that the crowd's wisdom has the potential for novice crowd usability inspection. Novice crowd usability inspection guided by a single expert's heuristic usability inspection is economical and time-efficient and a potential alternative method for usability evaluation for budget-constrained software organizations.

## REFERENCES

Abran, A., Khelifi, A., Suryn, W., and Seffah, A. (2003). Usability meanings and interpretations in iso standards. Software Quality Journal, 11(4):325–338.

Ambreen, T., & Ikram, N. (2016, August). A state-of-the-art of empirical literature of crowdsourcing in computing. In 2016 IEEE 11th International Conference on Global Software Engineering (ICGSE) (pp. 189-190). IEEE.

Ardito, C., Buono, P., Caivano, D., Costabile, M. F., Lanzilotti, R., Bruun, A., & Stage, J. (2011, January). Usability evaluation: a survey of software development organizations. In SEKE (pp. 282-287).

Bak, J. O., Nguyen, K., Risgaard, P., & Stage, J. (2008, October). Obstacles to usability evaluation in practice: a survey of software development organizations. In Proceedings of the 5th Nordic conference on Human-computer interaction: building bridges (pp. 23-32). ACM.

Botella, F., Alarcon, E., & Peñalver, A. (2014, September). How to classify to experts in usability evaluation. In Proceedings of the XV International Conference on Human Computer Interaction (pp. 1-4).

Botella, F., Alarcon, E., & Peñalver, A. (2013, November). A new proposal for improving heuristic evaluation reports performed by novice evaluators. In Proceedings of the 2013 Chilean Conference on Human-Computer Interaction (pp. 72-75)

Borys, M., & Laskowski, M. (2014, April). Expert vs Novice Evaluators. In Proceedings of the 16th International Conference on Enterprise Information Systems-Volume 3 (pp. 144-149). SCITEPRESS-Science and Technology Publications, Lda

Bruun, A. and Stage, J. (2015). New approaches to usability evaluation in software development: Barefoot and crowdsourcing. Journal of Systems and Software, 105:40–53.

Bruun, A., Gull, P., Hofmeister, L., & Stage, J. (2009, April). Let your users do the testing: a comparison of three remote asynchronous usability testing methods. In Proceedings of the SIGCHI Conference on Human Factors in Computing Systems (pp. 1619-1628).

Carletta, J. Assessing agreement on classification tasks: the kappa statistic. *Computational Linguistics*, 22(2) (1996), 249–254

Chattratichart, J., & Lindgaard, G. (2008). A comparative evaluation of heuristic-based usability inspection methods. In CHI'08 extended abstracts on Human factors in computing systems (pp. 2213-2220).

Cockton, G., & Lavery, D. (1999, August). A framework for usability problem extraction. In INTERACT (pp. 344-352).

De Lima Salgado, A., de Souza Santos, F., de Mattos Fortes, R. P., & Hung, P. C. (2018).

Guiding Usability Newcomers to Understand the Context of Use: Towards Models of Collaborative Heuristic Evaluation. In Behavior Engineering and Applications (pp. 149-168). Springer, Cham.

Desurvire, H. W., Kondziela, J. M., & Atwood, M. E. (1992). What is gained and lost when using evaluation methods other than empirical testing. People and computers, 89-89.

Estell' es-Arolas, E. and Gonz'alez-Ladr' on-De-Guevara, F. (2012). Towards an integrated crowdsourcing definition. Journal of Information science, 38(2):189–200.

Følstad, A., Anda, B. C., & Sjøberg, D. I. (2010). The usability inspection performance of work-domain experts: An empirical study. Interacting with Computers, 22(2), 75-87.

Garcia-Molina, H., Joglekar, M., Marcus, A., Parameswaran, A., & Verroios, V. (2016). Challenges in data crowdsourcing. IEEE Transactions on Knowledge and Data Engineering, 28(4), 901-911.

Ghaffar, H., Nasir, M. (2016). Usability heuristics for designing web dashboard. MS Thesis at International Islamic University, Islamabad, Pakistan.

Glassdoor (2021). Retrieved from https://www.glassdoor.com/Reviews/Employee-Review-uTest-RVW15073606.htm

Gomide, V. H., Valle, P. A., Ferreira, J. O., Barbosa, J. R., Da Rocha, A. F., & Barbosa, T. (2014). Affective crowdsourcing applied to usability testing. International Journal of Computer Scienceand Information Technologies, 5(1), 575-579.

Goodman, J. K., Cryder, C. E., and Cheema, A. (2013). Data collection in a flat world: The strengths and weaknesses of mechanical turk samples. Journal of Behavioral Decision Making, 26(3):213–224.

Gray, W. D., & Salzman, M. C. (1998). Damaged merchandise? A review of experiments that compare usability evaluation methods. Human-computer interaction, 13(3), 203-261.

Guaiani, F. and Muccini, H. (2015). Crowd and laboratory testing, can they co-exist? an exploratory study. In CrowdSourcing in Software Engineering (CSI-SE), 2015 IEEE/ACM 2nd International Workshop on, pages 32–37. IEEE.

Hasan, L., Morris, A., & Probets, S. (2012). A comparison of usability evaluation methods for evaluating e-commerce websites. Behaviour & Information Technology, 31(7), 707-737.

Häkli, A. N. N. A. "Introducing user-centered design in a small-size software development organization." Helsinki University of Technology, Helsinki (2005).

Hertzum, M., & Jacobsen, N. E. (2001). The evaluator effect: A chilling fact about usability evaluation methods. International journal of human-computer interaction, 13(4), 421-443.

Hollingsed, T., & Novick, D. G. (2007, October). Usability inspection methods after 15 years of research and practice. In Proceedings of the 25th annual ACM international conference on design of communication (pp. 249-255). ACM.

Hornbæk, K., & Frøkjær, E. (2008, April). Making use of business goals in usability evaluation: an experiment with novice evaluators. In Proceedings of the SIGCHI Conference on Human Factors in Computing Systems (pp. 903-912).

Howarth, J., Smith-Jackson, T., & Hartson, R. (2009). Supporting novice usability practitioners with usability engineering tools. International Journal of Human-Computer Studies, 67(6), 533-549.

Jeffries, R., Miller, J. R., Wharton, C., & Uyeda, K. (1991, April). User interface evaluation in the real world: a comparison of four techniques. In Proceedings of the SIGCHI conference on Human factors in computing systems (pp. 119-124). ACM.

Kittur, A., Chi, E. H., & Suh, B. (2008, April). Crowdsourcing user studies with Mechanical Turk. In Proceedings of the SIGCHI conference on human factors in computing systems (pp. 453-456). ACM.

Koutsabasis, P., Spyrou, T., Darzentas, J. S., & Darzentas, J. (2007). On the performance of novice evaluators in usability evaluations. Proc. PCI, Patra.

Kostkova, P., Garbin, S., Moser, J., and Pan, W. (2014). Integration and visualization public health dashboard: the medi+ board pilot project. In Proceedings of the 23rd International Conference on World Wide Web, pages 657–662. ACM.

Law, E. L. C., & Hvannberg, E. T. (2008, October). Consolidating usability problems with novice evaluators. In Proceedings of the 5th Nordic conference on Human-computer interaction: building bridges (pp. 495-498).

Lechner, B. and Fruhling, A. (2014). Towards public health dashboard design guidelines. In International Conference on HCI in Business, pages 49{59. Springer.

Liu, D., Bias, R. G., Lease, M., and Kuipers, R. (2012). Crowdsourcing for usability testing. Proceedings of the American Society for Information Science and Technology, 49(1):1–10.

Molich, R., & Dumas, J. S. (2008). Comparative usability evaluation (CUE-4). Behaviour & Information Technology, 27(3), 263-281.

Molich, R., Ede, M. R., Kaasgaard, K., & Karyukin, B. (2004). Comparative usability evaluation. Behaviour & Information Technology, 23(1), 65-74.

mTurk, Amazon Mechanical Turk. (2020). Retrieved from https://requester.mturk.com/tour.

Nielsen, J. (1989). Usability engineering at a discount. In Proceedings of the third international conference on human-computer interaction on Designing and using human-computer interfaces and knowledge-based systems (2nd ed.), pages 394–401. Elsevier Science Inc.

Nielsen, J. (1992, June). Finding usability problems through heuristic evaluation. In Proceedings of the SIGCHI conference on Human factors in computing systems (pp. 373-380).

Nielsen, J. (1994a). Usability inspection methods. In Conference companion on Human factors in computing systems (pp. 413-414).

Nielsen, J. (1994b). Enhancing the explanatory power of usability heuristics. In Proceedings of the SIGCHI conference on Human Factors in Computing Systems (pp. 152-158).

Peer, E., Vosgerau, J., and Acquisti, A. (2014). Reputation as a sufficient condition for data quality on amazon mechanical turk. Behavior research methods, 46(4):1023–1031.

Retelny, D., Robaszkiewicz, S., To, A., Lasecki, W. S., Patel, J., Rahmati, N., Doshi, T., Valentine, M., and Bernstein, M. S. (2014). Expert crowdsourcing with flash teams. In Proceedings of the 27th annual ACM symposium on User interface software and technology, pages 75–85. ACM.

Rosenbaum, S. (1989). Usability evaluations versus usability testing: When and why? IEEE transactions on professional communication, 32(4):210–216.

Sarı, A., Tosun, A., & Alptekin, G. I. (2019). A systematic literature review on crowdsourcing in software engineering. Journal of Systems and Software, 153, 200-219.

Shneiderman, B. (1996). The eyes have it: A task by data type taxonomy for information visualizations. In Visual Languages, 1996. Proceedings., IEEE Symposium on, pages 336{343. IEEE.

Stritch, J. M., Pedersen, M. J., & Taggart, G. (2017). The opportunities and limitations of using Mechanical Turk (Mturk) in public administration and management scholarship. International Public Management Journal, 20(3), 489-511.

Surowiecki, J. (2005). The wisdom of crowds. Anchor.

Upwork (2020). Retrieved from https://www.upwork.com/.

uTest (2020). Retrieved from http://www.utest.com/.

Wang, Y., Jia, X., Jin, Q., & Ma, J. (2017). Mobile crowdsourcing: framework, challenges, and solutions. Concurrency and Computation: Practice and experience, 29(3), e3789.

WHO-EU, D. (2016). WHO-EU Dashboard. Retrieved from http://www.euro.who.int/en/data-and-evidence/databases/european-health-for-all-database-hfa-db.

Xu, A., Rao, H., Dow, S. P., & Bailey, B. P. (2015, February). A classroom study of using crowd feedback in the iterative design process. In Proceedings of the 18th ACM conference on computer supported cooperative work & social computing (pp. 1637-1648).

Yen, P. Y., & Bakken, S. (2009). A comparison of usability evaluation methods: heuristic evaluation versus end-user think-aloud protocol–an example from a web-based communication tool for nurse scheduling. In AMIA annual symposium proceedings (Vol. 2009, p. 714). American Medical Informatics Association.

Yuan, A., Luther, K., Krause, M., Vennix, S. I., Dow, S. P., & Hartmann, B. (2016, February). Almost an expert: The effects of rubrics and expertise on perceived value of crowdsourced design critiques. In Proceedings of the 19th ACM Conference on Computer-Supported Cooperative Work & Social Computing (pp. 1005-1017).

Zhao, Y., & Zhu, Q. (2014). Evaluation on crowdsourcing research: Current status and future direction. Information Systems Frontiers, 16(3), 417-434.

## Appendix A: Web-Links for Test Objects

1. WHO-EU Web Dashboard
   [Weblink]: https://gateway.euro.who.int/en/
2. [Weblink]: www.ebay.com
3. [Weblink]: www.booking.com

## Appendix B: Usability Heuristics for web dashboard

Heuristics for Web Dashboard & Information Visualization [Web-Link]: https://drive.google.com/file/d/1vYkaKHegMLqUHlk4KklTImlwnvUzYyI6/view?usp=sharing

## Appendix C: Questionnaire for Crowd Usability Evaluation

1. Questionnaire for WHO-EU Web Dashboard [Web-Link]: https://docs.google.com/a/iiu.edu.pk/forms/d/1I9-KCINkMWrOPhtwxnzGBr6TOErb-IQeH0u390IX_60/viewform
2. Questionnaire for www.ebay.com [Web-Link]: https://drive.google.com/file/d/1z5lLGCn_FPeVlgMgXw3i8s5ptrK00aSn/view?usp=sharing
3. Questionnaire for www.booking.com [Web-Link]:https://drive.google.com/file/d/1Xp-FqP5E_yCUajkhlg1BhYQxYwOTPCpg/view?usp=sharing

## Appendix D: Key Usability Issues

https://drive.google.com/file/d/13nwNSdVi7sdxNHDsY3bCpwb4jMzcyGt7/view?usp=sharing

## Appendix E – A Coding Scheme for Usability Problem Detection & Classification

| Code | Short Description | Definition |
|---|---|---|
| MP | Minor problem | Causes a few seconds delay to extract desired information or understanding it |
| SP | Serious problem | Causes 1-5 minutes delay and prevents a user from understanding or extracting desired information, but the user somehow manages the information flow |
| CP | Critical problem | Prevents the user from obtaining desired information or understanding it, leading the user to complete failure |
| FP | False-positive | A usability issue is considered false positive, either if it is not reproducible or fixing that would not improve the usability of the system |
| CM | Comment | A comment can be a problem, or a bug reported by a usability evaluator |
| P-CM | Problem comment | A comment that is classified as a minor, serious, or critical problem |
| A-CM | Atomic comment | A comment that cannot be further divided into more comments. Atomic comments can be resolved without affecting other usability issues |
| I-CM | Identical comment | Two or more comments are identical if fixing one comment will fix other comments as well |
| IS | Issue | A set of one or more identical atomic comments |
| K-IS | Key issue | An issue that is reported by three experts or two experts and at least one crowd usability inspector, and it is, on average, marked as important, i.e., serious or critical problem. |
| C-CL | Contradictory classifications | This occurs when the same underlying usability finding is classified differently by more than one usability evaluator. For example, one expert may classify the finding as a serious usability problem while others may mark it as a minor usability problem. |

## Annex A: Usability Heuristics by Jakob Nielsen

10 Usability Heuristics by Jakob Nielsen for User Interface Design [Web-Link]: https://drive.google.com/file/d/1A6h0hTGUeGatH1XHynHMG29UI5ezmrxq/view?usp=sharing